\documentclass[preprint, 5p, twocolumn]{elsarticle}
\journal{Physics Letters A}
\usepackage[colorlinks=true, citecolor=red, urlcolor=blue ]{hyperref}
\usepackage{epsfig,newlfont,bm,subfigure,palatino,mathtools,braket,times,soul,setstack}
\usepackage{color}
\usepackage[utf8]{inputenc}
\usepackage[sc,osf]{mathpazo}
\usepackage[normalem]{ulem}
\usepackage[T1]{fontenc}
\usepackage{multirow}
\usepackage{ulem}
\usepackage{graphics}
\usepackage{graphicx}
\definecolor{indiagreen}{rgb}{0.07, 0.53, 0.03}
\newcommand{\tm}[1]{{\color{indiagreen}#1}}

\begin{document}

	\title{
		Quantum illumination with noisy probes: Conditional advantages of non-Gaussianity
	}

	\author{Rivu Gupta\(^1\), Saptarshi Roy\(^2\), Tamoghna Das\(^3\), and Aditi Sen(De)\(^{1}\)}
	
	\address{\(^1\)Harish-Chandra Research Institute, HBNI, Chhatnag Road, Jhunsi, Allahabad 211 019, India}
	\address{\(^2\) QICI Quantum Information and Computation Initiative, Department of Computer Science, The University of Hong Kong, Poklufam Road, Hong Kong}
        \address{$^3$ Department of Physics, Indian Institute of Technology Kharagpur, Kharagpur-721302, India}
	
	\begin{abstract}
		Entangled states, like the two-mode squeezed vacuum state, are known to give quantum advantage in the illumination protocol, a method to detect a weakly reflecting target submerged in a thermal background. We use non-Gaussian photon-added and -subtracted states, \textcolor{black}{affected by local Gaussian noise on top of the omnipresent thermal noise, as probes in the illumination protocol}. Based on the difference between the Chernoff bounds obtained with the coherent state and the non-Gaussian state having equal signal strengths, whose positive values denote quantum advantage in illumination, we highlight the hierarchy among non-Gaussian states, which is compatible with correlations per unit signal strength, \textcolor{black}{although the Gaussian states offer the best performance}.  Interestingly, such hierarchy is different when comparisons are made using the Chernoff bounds. The entire analysis is performed in the presence of different imperfect apparatus like faulty twin-beam generator, imperfect photon addition (subtraction) as well as with noisy non-Gaussian probe states.
	\end{abstract}


	\maketitle
	
	\section{Introduction}
	\label{sec:intro}

	The non-classical features offered by quantum mechanics have revolutionized the development of modern technologies, ranging from computation \cite{Feyn,QComp,QComp2,QComp3,QComp4}, and communication \cite{AditiComm,GisinComm,AditiComm2,AditiComm3,Comm4,Comm5,Comm6} to metrology and memory devices \cite{metrologymemory}, far superior to their classical counterparts.  
	Remarkable protocols like quantum teleportation \cite{BBCJPW,TF2,Pirandola,Pawel,Fedorov,Seida,Badziag,Verstraete}, dense coding \cite{bennettwiesner,DC2,DCCamader,Bruss,Horo12,Aditicap,DCCTamo1,Tamoghna}, and quantum key distribution \cite{BB84,EkertCrypto,Crypto3,Crypto4,Keydist} boost the communication sector both with and without security while in the computational domain,  algorithms based on quantum mechanics were discovered which offer as much as exponential speedup compared to the known methods in a classical computer  \cite{shor2, dz, grover}.
	The enhanced performance in most of the schemes relies on the amount of quantum correlations (QCs) present in the system, establishing them as the resource \cite{rt} for quantum advantage.
	
\begin{figure}[t]
		\centering
		\includegraphics[width=\linewidth]{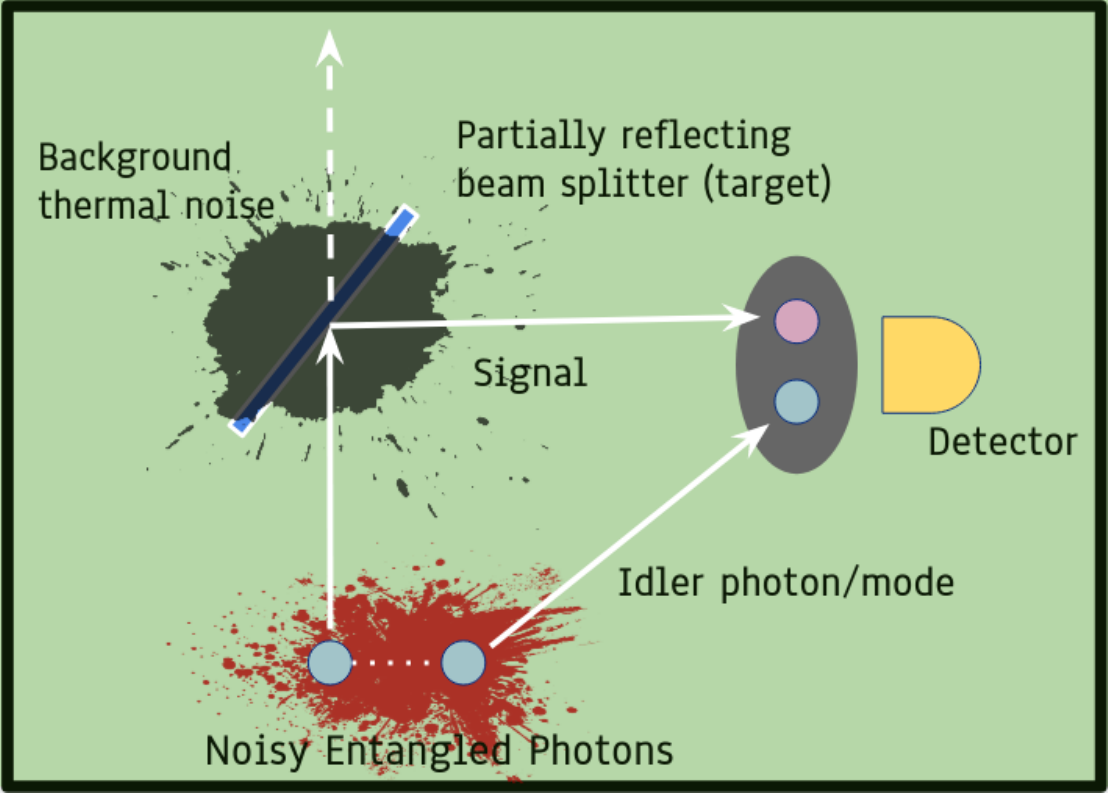}
		\caption{ A schematic representation of quantum illumination with noisy probes. The signal mode is sent towards the target, which is modeled by a weakly reflecting beam splitter. The idler mode is stored till the reflected signal comes back and the measurement is performed jointly on the two modes, to infer the presence or absence of the target. A strong thermal background is always present, neighboring the target. Moreover,  we consider different noise models in the probe states.}
		\label{fig:schematic}
	\end{figure}

	In the field of quantum metrology \cite{Metro1,Metro2,Metro3,Metro4,Metro5}, illumination is the process of detecting a target with low reflectivity encapsulated in a noisy thermal background \cite{GIllu,GIllu4,Illu2}. In particular, a probe signal is sent toward the target, and its presence or absence is inferred by analyzing the reflected beam.
	In quantum illumination (QI) \cite{Lloyd,Shapiro-rev,TrepsArxiv,Illu22,Illu3,Illu4,Illu5,Illu6,Illu7, Nair_Optica_2020, Shi_arXiv_2022}, it was shown that the sensing capabilities, for a target modeled by a beam splitter (BS), can be improved by using entangled probes like the two-mode squeezed vacuum (TMSV) state  \cite{GIllu6,GIllu7,GIllu8,AsymmSqueeze,GIllu10}. In this situation, one mode of the entangled pair is used as the signal mode, while the other mode (acting as idler) is directly sent to the detector to be stored and measured jointly after the signal mode reflected from the beam splitter returns (see Fig. \ref{fig:schematic} for a schematic of the protocol).  
	In this context, it was also shown that the initial shared entangled state comprising the signal and the idler modes still remains beneficial even in the presence of loss and noise, which can destroy the resource \cite{NonEntAdv,EntGone,EntReact}.
	Moreover, the idler mode needs to be stored until the time the signal returns, which significantly reduces the range of QI since the storage of the idler for a longer duration is difficult \cite{IdlerStorage}.  
	For the classical illumination protocol using coherent states, homodyne detection turns out to be the optimal one \cite{DoubleHomodyne} while
	for the Gaussian QI, more involved detection schemes are required to extract the quantum advantage which has been predicted theoretically \cite{Homo,GIllu3,GIllu2,GIllu5,GIllu9,Neyman-Pearson,HardDetection,Detect,Detect2}.
	Finally, unlike other quantum  devices, it is not yet clear whether quantum correlations, especially the entanglement content of the initial Gaussian state are responsible for quantum advantage in the illumination process
	\cite{MoreThanEnt,MoreThanEnt2,MoreThanEnt3,MoreThanEnt4,MoreThanEnt5,MoreThanEnt6,AsymmSqueeze}. 

	On the other hand,  it has been shown that non-Gaussian states, created by adding (subtracting) photons in (from) the TMSV state,  which possess a higher amount of QCs than that of the parent Gaussian state, 
	have the potential to provide an advantage in the  performance of QI \cite{NGIllu,NGIllu2,NGIllu3,NGIllu4} in a noiseless 
	situation \cite{GIllu3}. In this paper, we investigate the efficiency of non-Gaussian probe states in the presence of different kinds of noise and imperfections. The performance of QI is typically quantified by the minimum error probability for distinguishing the presence or absence  of target states
	which is upper bounded by the quantum Chernoff bound (CB)  \cite{Chernoff,Chernoff2,Chernoff3,Bound1,Discriminate2,Bhattacharyya,ChernoffTight}. In a noiseless scenario, \textcolor{black}{i.e., only in the presence of the thermal background}, we report a monotonic decrease of the CB with an increasing number of added or subtracted photons from a single mode of the two-mode squeezed vacuum state. Symmetric two-mode operations (i.e., when an equal number of photons are added or subtracted from both modes)  lower the CB even further from the single-mode value. 
	We also observe that the non-Gaussian states can lower the CB even when the target reflectivity is very small and such an advantage increases with the number of photons added (subtracted). 
	\textcolor{black}{In a situation when noise, apart from the inevitable thermal background, acts locally on the individual modes,} we find that the robustness in the CB against Gaussian noise, which is mixed with the non-Gaussian states, increases with the increase of non-Gaussianity. Similar trends can also be exhibited when the twin beam generator producing the parent TMSV state is faulty or when there is an imperfection in the photon addition and subtraction processes. According to the low values of Chernoff bounds, we can provide a hierarchy among the photon-added and -subtracted states which is quite similar to the QC present in these states (cf. \cite{NGIllu, NGIllu2}).
	
	We now refer to a state to be quantum advantageous in the illumination protocol if the difference between the Chernoff bound of a coherent state and that of a given non-Gaussian state having the same signal strength, is strictly positive. 
	Based on it, we now classify different non-Gaussian states according to their performance in QI, \textcolor{black}{even though such states cannot surpass the TMSV state having the same signal strength}.     
	We report that for low reflectivity, photon addition in the idler mode (or photon subtraction in the signal mode) and photon subtraction from both the modes do yield quantum advantage while photon addition in both the modes or only in the signal mode cannot beat the classical limit, thereby giving a non-positive difference with the coherent state CB. However, the hierarchy changes with an increase in the reflectivity of the BS. In a noisy scenario where the signal transmission line is assumed to be noisy, \textcolor{black}{over and above the thermal background}, thereby affecting both non-Gaussian as well as coherent states, we interestingly observe that quantum advantage with non-Gaussian states increases even with the increase of noise up to a certain threshold value and then decreases as expected. Additionally,
	we arrive at a minimum operational efficiency of the photon addition (subtraction) apparatus to obtain quantum advantage, which also takes care of the probabilistic nature involved in the generation of non-Gaussian states from the TMSV states. This analysis also helps us to decide the most favorable resource, non-Gaussian or Gaussian,  depending on the apparatus available. Note that although the entirety of the work focuses on the advantage rendered by just a single copy of the probe state,  the extension to QI with multiple copies is quite straightforward in our framework. 
	Moreover, we report that the ranking of non-Gaussian states according to the advantage in the QI turns out to be in good agreement with correlations, quantified by mutual information and entanglement of the given state, per signal photon.

	The paper is organized in the following way. In Sec.\;\ref{sec:pre},  we provide the prerequisites which include the Chernoff bound (the upper bound on the efficiency of the illumination protocol), its classical limit, and the non-Gaussian states together with the noise models which we will use in our calculations. This is followed by Sec.\;\ref{sec:ng} where we elucidate the advantages offered by non-Gaussian states, with a particular focus on the comparison between the single-mode addition and subtraction of photons and the two-mode operations. We then move on to the definition of \textit{quantum advantage} and show how only certain non-Gaussian states can actually outperform the classical protocol, while others fail to do so. In Sec.\;\ref{sec:noisy}, we introduce noise in probe states, modeled by Gaussian local noise and faulty twin beam generators, and establish the robustness exhibited by non-Gaussian states to various noise models while in Sec. \ref{sec:advantage_ng}, we compare Gaussian TMSV states with non-Gaussian states in two ways -- one when non-Gaussian apparatus is inefficient and another via the correlation content of the states. 
	We end our paper with the discussions of results in Sec.\;\ref{sec:conclusion}.
	
	\section{Ingredients for analyzing quantum illumination with non-Gaussian resources}	
	\label{sec:pre}
	
	In this section, we discuss the tools required to analyze QI using non-Gaussian states both in the presence and absence of noise. 
	We describe the various components involved in the QI protocol. In particular,  we specify the methodology to compute the performance of the QI protocol including numerical recipes used for evaluation. We then move on to present the various noise models that we employ to investigate noisy QI. \textcolor{black}{A brief primer about the non-Gaussian states to be used as probes for QI generated by photon addition and subtraction is discussed in Sec. $1$ of the Supplementary material}. 
	

	

	\subsection{Elements of quantum illumination}

	The QI protocol comprises three main components -- $(i)$ the probe (the signal and the idler), $(ii)$ the weakly reflecting target embedded in a thermal background, and $(iii)$ the detection scheme involving a joint measurement of the signal and the idler,  
	as depicted schematically in Fig. \ref{fig:schematic}. The task of inferring the presence or absence of the target reduces to a two-state discrimination problem \cite{StateDiscri,Discriminate,Discriminate2,Discriminate3,Discriminate4,Discriminate5,Discriminate6,Discriminate7,Discriminate8,Disscriminate9,Discriminate10,Helstrom} and hence, the performance of QI relies on the efficiency of the distinguishability protocol. 
	The figure of merit of QI is then the minimum error probability that arises while discriminating the two non-orthogonal states, each of which corresponds to either of the two hypotheses: $H_0 $ representing the scenario when the target is absent while $H_1 $ identifying the presence of the target. 
	
	Mathematically, the weakly reflecting target is modeled by a beam splitter with low reflectivity, $\kappa$. It is immersed in a thermal bath,  $\rho_{T}$, with mean photon number, $N_B$. The entangled probe used for QI is $\rho_{IS}$, where the subscript, $I$, is for the idler and $S$ represents the signal mode.  
	When the target is present (hypothesis $H_1$),  the reflected signal is admixed with the thermal noise, and the resultant state in the detector is given by
	\begin{eqnarray}
	\hspace{-1em} \text{Target present}\,  (H_1):	~\rho_1 = \text{Tr}_T (\hat{U}_{ST}\, \rho_{IS}\otimes \rho_T \,  \hat{U}_{ST}^{\dagger}),
	\label{rho_1_exp}
	\end{eqnarray}
	where $\hat{U}_{ST}$ is the unitary representation of the BS  acting on the signal and the thermal state,  having the form $\hat{U}_{ST}(\xi)=\exp(\xi \hat{a}_S^{\dagger} \hat{a}_T - \xi^* \hat{a}_S \hat{a}_T^{\dagger})$, with $\xi = \sin^{-1}\sqrt{\kappa}$. Here  $\hat{a}_S,~\hat{a}_I$, and $\hat{a}_T$  are the annihilation operators for the signal, the idler, and the thermal modes respectively.
	For a very generic Gaussian or non-Gaussian state, $\rho_{IS}$, 
	Eq. (\ref{rho_1_exp}) reduces to
	\begin{eqnarray}
	&& \hspace{-1em} \rho_1 = \sum_{n,n' = n_0}^{\infty} \sum_{m = 0}^\infty \sum_{r = 0}^{n \pm l}\sum_{s,s' = 0}^{m}
	f\left(\{n\}, \{r\},k,l,\kappa,N_B\right) \times
	\nonumber  \\  
	&& \hspace{-3em} \nonumber
	\ket{n \pm k, n + m \pm l - r - s} \bra{n' \pm k, n' + m \pm l - r - s}_{IS}, \\
	\label{rho1}
	\end{eqnarray} 
	where the set $\{n\} = (n,n',m)$, $\{r\} = (r,s,s')$ and $n_0 = 0$ for addition and $n_0 = \max\{k,l\}$ for photon subtraction.
	The prefactor $f\left(\{n\}, \{r\},k,l,\kappa,N_B\right)$ is given by
	\begin{eqnarray} 
	\nonumber && \hspace{-2em} f = \frac{c_n^{\pm k, \pm l} c_{n'}^{\pm k,\pm l}}{\sqrt{(n \pm l)!(n' \pm l)!}} \frac{(N_B)^{m}}{(1 + N_B)^{1 + m}} \frac{1}{m!} \binom{m}{s}\binom{m}{s'} (r + s)!  \\ \nonumber && \hspace{-2em} \binom{n \pm l}{r} \binom{n' \pm l}{r + s - s'} (-1)^{n + m -s +s'} \kappa^{\frac 12 (n + n') - r + s' \pm l} \nonumber \\ &&  \nonumber \hspace{-2em} (1-\kappa)^{m + r - s'}   \sqrt{(n  + m - r - s \pm l)!(n' + m - r  - s \pm l)!} \\ 
	\label{prefactor}
	\end{eqnarray} 
	where, for the TMSV state, $k = l = 0$ and $\pm l ~ \text{and} ~ \pm k$ represent photon addition and subtraction from the signal and idler modes respectively.
	The coefficient, $c_n^{\pm k, \pm l}$, for various combinations of photonic operations are given in Sec. $1$ of the Supplementary material. 
	
	Let us move to the hypothesis $H_0$, i.e., when the signal is lost and the detector just gets the thermal state and the idler. In this case, the state  simply takes the form as
	\begin{equation}
	\text{Target absent} \, (H_0): ~\rho_0 = \text{Tr}_S ~\rho_{IS} \otimes \rho_{T}.
	\label{rho_0_exp}
	\end{equation}
	The explicit expression for $\rho_0$ reads as
	\begin{eqnarray}
	\nonumber && \rho_0 = \sum_{n = n_0}^{\infty} \sum_{m = 0}^{\infty}  (c_n^{(\pm k,\pm l)})^2  \frac{N_B^{m}}{(1 + N_B)^{m+1}} \times \\
        && ~~~~~~~~~~~~~~~~~~~~~ \ket{n \pm k,m}\bra{n \pm k ,m}_{IT}, 
	\label{rho0}
	\end{eqnarray}
	where the lower limit $n_0$ of the first summation satisfies the same condition as for $\rho_1$ in Eq. \eqref{rho1} and the subscripts $I$ and $T$  denote the idler and the thermal states respectively.
	
	As mentioned before, the efficiency of QI reduces to the problem of effectively distinguishing  $\rho_0$ from $\rho_1$, or multiple copies (say 
 $M$) of them, with the least possible error using an optimal measurement scheme. Assuming that there is a priori an equal probability of the target being present or absent,
	the minimum error probability for distinguishing $\rho_0^{\otimes M}$ and $\rho_1^{\otimes M}$  can  then be  expressed as \cite{GIllu}
	\begin{eqnarray}
	P_M = \frac{1}{2}\left(1 - \frac{1}{2}||\rho_0^{\otimes M} - \rho_1^{\otimes M}||_1\right),
	\end{eqnarray}
	where $||X||_1 = \text{tr} |X|$ represents the trace norm \cite{Hell, NGIllu,NGIllu2, TrepsArxiv}. 
	From the above expression, it is not easy to calculate the error probability, and hence we will focus on \textcolor{black}{an} upper bound of it, given by  
        \textcolor{black}{
	\begin{eqnarray}
	P_M \leq \mathcal{Q}_M = \frac{1}{2} \Big(\min_{0 \leq \alpha \leq 1} \text{tr}~[\rho_0^\alpha \rho_1^{1-\alpha}]\Big)^M.
	\label{eq:CBdef}
	\end{eqnarray}
	where the $\mathcal{Q}_M$ is known as the quantum Chernoff bound  \cite{Chernoff,Chernoff2,Chernoff3,Bound1,Discriminate2,Bhattacharyya,ChernoffTight}. Note that twice the Chernoff bound for $M$ probe states is just the $M$-th power of the same for a single copy, i.e.,
 \begin{equation}
 2 \mathcal{Q}_M = 2^M \mathcal{Q}^M,
 \label{chernoff_M}
 \end{equation}
 where $\mathcal{Q} = \dfrac{1}{2} \min_{0 \leq \alpha \leq 1} \text{tr}~[\rho_0^\alpha \rho_1^{1-\alpha}]$ represents the Chernoff bound for a single copy of the probe state.} 
	Moreover, since the quantum CB is asymptotically tight \cite{ChernoffTight,Discriminate2},  the  bound for one copy of the probe state dictates the hierarchies of error probabilities obtained in the large $M$ case.
	We, therefore, throughout the manuscript,  work with the single shot case, i.e., for $M =1$. Results for any finite $M$ can simply be obtained from Eq. \eqref{chernoff_M}. \textcolor{black}{In Sec. $2$ of the Supplementary material, we discuss the classical limit of the Chernoff bound, required to determine the quantum advantage obtained while using non-Gaussian probes.}
	
	\textcolor{black}{\textbf{Remark}. The Chernoff bound obtained for $M=1$ is not a good measure of the performance of quantum illumination. The bound becomes good only for large $M$ where it becomes asymptotically tight. However, mathematically, the discrimination error probability for $M$ copies of the probe state is bounded by $P_M \leq \frac{1}{2} \mathcal{Q}^M$.  Since the Chernoff bound scales as the exponent of $\mathcal{Q}$, it is sufficient to investigate only $\mathcal{Q}$ (for $M = 1$) as also done in Ref. \cite{NGIllu}. This very fact makes the Chernoff bound computable for large $M$. Moreover, it also implies that when the performance of two different probe states is compared with respect to the target detection in quantum illumination, the one with a lower $\mathcal{Q}$ is better. Our numerical methodology to compute the Chernoff bound is thoroughly discussed in Sec. $3$ of the Supplementary material}.

	

	\subsection{Noise models for probes}
	\label{subsec:noise}
	
	To analyze the case of noisy illumination protocol, we introduce certain types of imperfections in the probe states. \textcolor{black}{Note that, the background thermal noise is inherent in the illumination protocol since the target itself is immersed in it. However, in most of the QI literature, the signal and idler are assumed to be free from any noise, i.e., the probe states are considered to be pure. Such an assumption is too idealistic, and, therefore, we consider additional imperfections that may affect the protocol. We examine noise models that disturb both the signal and the idler modes through imperfections in the resource generation protocol. Furthermore, local noise models affecting the signal mode during transmission are also considered separately. Note that another significant hurdle in the illumination scheme is the problem of idler storage, which, however, is fundamentally different from the losses that we wish to address in the manuscript, and we skip this investigation in this work. Let us now enlist the noise models necessary for the analysis of the illumination protocol involving imperfect resource states.}
	
	\subsubsection{Local noise in probes}
	\label{subsubsec:LN}
	
	We consider two-mode entangled states in the presence of local noise, acting independently on each mode. The noise model which we consider yields the  mixed state (see ref. \cite{SappyTamoGBell}), given by
	\begin{equation}
	\rho = (1 - p)|\psi \rangle \langle \psi| + p \left(\sum_{n = 0}^{\infty}\mu_n |n \rangle \langle n | \otimes \sum_{m = 0}^{\infty} \nu_m |m \rangle \langle m|\right)
	\label{noisy_state}
	\end{equation}
	where $|\psi \rangle$ denotes the non-Gaussian entangled state, and $\sum_{n = 0}^{\infty} \mu_n = \sum_{m = 0}^{\infty} \nu_m = 1$. In our analysis, $\mu_n$ and $\nu_m$ are taken to have a Gaussian form, with
	\begin{eqnarray}
	\nonumber && \mu_n  =  \frac{2}{1 + \vartheta_3(0,e^{-\sigma_1^{-2}})}e^{-n^2/ \sigma_1^{2}}, \\
	\mbox{and} ~~~~\, \, \, && \nu_m = \frac{2}{1 + \vartheta_3(0,e^{-\sigma_2^{-2}})}e^{-m^2/ \sigma_2^{2}}.
	\label{gaunoise}
	\end{eqnarray}
	Here, $\sigma_1$ and $\sigma_2$ are the chosen noise parameters that control the average number of photons in either mode of the noise part, and $\vartheta_3$ is the Jacobi theta function \cite{jt} of order 3. Throughout our analysis, we set $\sigma_1=\sigma_2 = 1$. The results remain qualitatively similar with other choices of \(\sigma_i\), as long as they are not too high to erase the quantum advantage.\\
	\textbf{Note 1.} An exemplary Gaussian noise model can be thought to be analogous to probabilistically mixing white noise to a state in the finite-dimensional formalism (eg. the Werner state). The coefficients are chosen to be Gaussian, since several noise models can be approximated by the same and it is easier to handle both theoretically and experimentally, thereby being applicable in quantum information processing tasks like continuous-variable quantum key distribution,\cite{noiseex3,noiseex2,noiseex4, noiseex5, noiseex6, noiseex1, noiseex7, noiseex8, noiseex9, noiseex10, noiseex11}). We also find that qualitatively similar results can be obtained, if, instead of the coefficients coming from a Gaussian or more specifically half-Gaussian distribution, the local states that get mixed with the initial states are thermal, i.e., 
\begin{eqnarray}
    \mu_n = (1 - e^{-\beta_1})e^{-\beta_1 n }, \\
    \nu_m = (1 - e^{-\beta_2})e^{-\beta_2 m}.
\end{eqnarray}

	\subsubsection{Faulty generation of two-mode state}
	\label{subsec:faulty}
	
	Another inefficiency during the production of probes can occur when the twin beam generator making the TMSV state is faulty \cite{SappyTamoGBell}. As a result, the probe state has less squeezing than actually expected. Therefore, the protocol may be designed for a squeezing parameter $r$ but the actual resource may have less efficiency due to a lower squeezing parameter $r' \, (<r)$ which in turn translates to $x' \, (<x)$. Thus the performance obtained may not be optimal and there may even be a situation when there is no  advantage at all, over the blind guess. We explore the role of non-Gaussianity in this respect, to figure out whether quantum advantage can be increased or even restored through the application of photon-added or -subtracted states in this scenario.
	
	\section{Non-Gaussian quantum illumination}
	\label{sec:ng}
	
	It has already been established that the introduction of  non-Gaussianity in the resource state through photon addition and subtraction can generate a high amount of quantum correlations.  Hence there is a possibility that these classes of states can lead to a lower error probability, in terms of CB in the illumination protocol, than that of the coherent state. 
	The  indication in this direction was obtained  by taking examples of  photon-added and -subtracted states \cite{NGIllu,NGIllu2}.
	
	We here analyze the performance  of QI via computing the Chernoff bound into two situations -- (1) single mode operations which include adding  photons in  either the idler or the signal mode; (2) when addition and subtraction are performed in both the modes. Moreover, we study the effects of variation of  target reflectivity on the performance of quantum illumination. \\

	\subsection{Single mode operations: Signal vs. Idler}

\begin{figure}
		\centering
		\includegraphics[width=\linewidth]{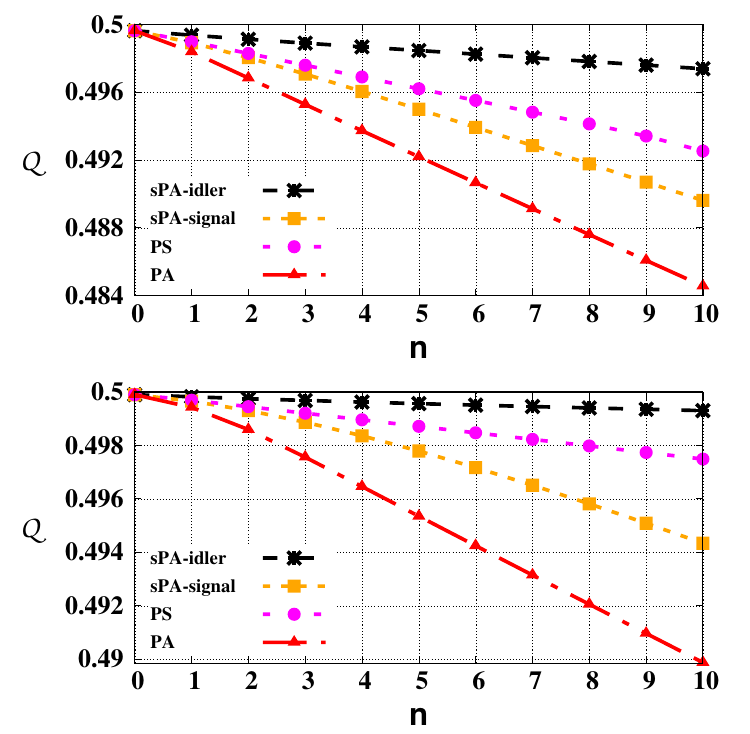}
		\caption{(Color Online.) Chernoff bound, \(\mathcal{Q}\), (ordinate) against the number of photons added (subtracted), \(n\), (abscissa) to create different non-Gaussian states from TMSV.  
			Stars and squares are for 
			the states with photons added either in the idler mode, denoted by sPA-idler, and signal mode (sPA-signal) respectively while 
			circles and triangles represent states when  \(n\) number of photons are added (PA) and 
			subtracted (PS) from both modes. Notice that \(\mathcal{Q}\) for sPA-idler and sPA-signal are equal to that of photon-subtracted state from the signal mode and idler modes respectively. \textcolor{black}{Here the target reflectivity, $\kappa = 0.01$, and the mean background photon number, $N_B = 1.0$.} The squeezing parameter is set to \textcolor{black}{ $x = 0.2 ~ (\approx 4 dB)$ for the upper panel and $x = 0.05 ~ (\approx 2 dB)$} for the lower panel. All the axes and dimensionless. }
		\label{cbound_double}
	\end{figure}

	Let us  add  photons in \textcolor{black}{a} single mode of the TMSV and compute the QI performance. As mentioned before, the QI scheme with subtracting photons from a single mode is already included in the results for photon addition. For example, we know  $|\psi_r^{(k,0)}\rangle = |\psi_r^{(0,-k)}\rangle$. Let us specify the behavior of Chernoff bound, \(\mathcal{Q}\), under single-mode operations. 
	\begin{enumerate}
		\item \textit{Monotonicity.} \(\mathcal{Q}\)  decreases monotonically with the number of added photons, \(n\), both in the idler and the signal modes as shown in Fig. \ref{cbound_double} 
		with \textcolor{black}{$x = 0.05$ $(\approx 2 dB)$ and $x=0.2 ~ (\approx 4 dB)$}. Interestingly, owing  to the low squeezing at $x = 0.05$,   the error probability for the TMSV state cannot go below \(0.5\) although non-Gaussian states are successful in doing so. However, later we will address  whether such decrements really imply the advantage in QI or not. 
		
		\item \textit{Asymmetry.} Although 
		the CB behaves monotonically with the added number of photons in a single mode,   it depends on the mode, (signal or idler), in which the photons are added. In particular, photons added in the idler mode always give a poorer detection probability than the case when photons are added in the signal mode of the entangled state. 
		The effect becomes prominent with the addition of a higher number of photons, thereby inducing more non-Gaussianity in the state, see Fig. \ref{cbound_double}. \textcolor{black}{Moreover, we observe that when photons are added in both the modes of the TMSV, we get a lower CB compared to the single mode operations, with the symmetric combination yielding the best performance. Notice, however, that among all the non-Gaussian states, the states with photons added in both modes can give the lowest  CB, although it does not lead to an advantage in QI, as we will argue later.} \\
	\end{enumerate}
	\textbf{Note $\mathbf{1}$.} 	For measures like entanglement ($E$), there is a symmetry between the modes, and we have $E(|\psi_r^{(k,0)}\rangle) = E(|\psi_r^{(0,k)}\rangle)$. However, for CB, in case of single-mode photonic operations, $\mathcal{Q}(|\psi_r^{(k,0)}\rangle) < \mathcal{Q}(|\psi_r^{(0,k)}\rangle)$ and hence there is an asymmetry inherent in its definition. Therefore,  
	for photon subtraction, we get a reversed relation compared to that of addition, i.e.,  subtraction of photons from the idler mode induces lower CB than that of the photon subtraction from the signal mode, i.e., $\mathcal{Q}(|\psi_r^{(-k,0)}\rangle) > \mathcal{Q}(|\psi_r^{(0,-k)}\rangle)$. 

	
	
	\textbf{Note $\mathbf{2}$.} \textcolor{black}{All the following analysis and figures correspond to the effect on the Chernoff bound, $\mathcal{Q}$, or the quantum advantage, $\Delta$ (defined in the next subsection) due to the addition and subtraction of photons, which we take up to $n = 10$. It is known that addition or subtraction of such a large number of photons is experimentally very hard to achieve, thereby giving us the motivation to  consider the probabilistic activation of QI in Sec. \ref{sec:advantage_ng}. We  perform such an analysis as a theoretical exercise to demonstrate how the Chernoff bound is affected by adding (subtracting) a large number of photons and to obtain the hierarchies among non-Gaussian states.}





	\subsection{Does reduction in Chernoff bound ensure quantum advantage in QI?}

	The analysis up to now shows the benefit of non-Gaussianity in terms of a ubiquitous decrease in CB when more photons are added or subtracted from the TMSV. 
	If  CB reduction does translate into quantum advantage, then one can conclude that the non-Gaussian photon-added and -subtracted states show better performance than the parent TMSV (cf. \cite{NGIllu, NGIllu2}).
	Note that such a comparison scheme may be considered natural and intuitive since it follows the same strategy when considering enhancements of entanglement \cite{Cerfaddsub, tamo2016}, violations of Bell inequalities \cite{SappyTamoGBell} etc. 
	
	Moreover, we know that photon addition and subtraction can be considered as a sort of distillation procedure  where fewer higher entangled states are obtained from a large number of low entangled states, and hence for entanglement, content, such  comparison  is perfect.
	However, for  schemes like quantum illumination where a higher number of copies are involved, the reduction of copies via distillation must be associated with the performance-calibration scheme, which we will  consider in a subsequent section. Hence comparing an identical number of copies of Gaussian and non-Gaussian states does not lead to a fair conclusion. If they are done, they must be considered independent states and the comparison must be made by fixing  one relevant physical quantity, which, in the case of QI, is the  signal strength \cite{NGIllu}.
	
	Let us set up a more appropriate scheme to assess the quantum advantage for QI. 
	Under the more realistic method of comparison, the hierarchy of states gets altered, and sometimes, the non-Gaussian states which provided 
	low CB fail to provide any quantum advantage. Therefore, the central question that we  want to address, under the new comparison methods, is \textit{when does non-Gaussianity provide a quantum advantage in illumination?}
	
	The quantum advantage in QI for the TMSV state is  defined by the difference between the Chernoff bounds for the TMSV state and  that of the coherent state with identical intensity in the signal mode. 
	In a similar fashion,  \textit{quantum advantage for the photon-added or -subtracted states} can be defined  as
	the difference between the Chernoff bounds for the non-Gaussian states and the coherent state  for fixed signal strength. In this case, for a given photon-added  (-subtracted) state, $|\psi_r^{(\pm k, \pm l)}\rangle$, we first compute the average number of photons it possesses in its signal mode, which reads as 
	\begin{eqnarray}
	N_S = \sum_{n =\frac{k - (\pm k)}{2}}^\infty (n\pm k) ~|c_n^{(\pm k, \pm l)}|^2.
	\label{eq:ns1}
	\end{eqnarray}
	The classical limit is then found by using the coherent state which has the same  signal strength  as in $|\psi_r^{(\pm k, \pm l)}\rangle$, i.e.,  $N_S(|\psi_r^{(\pm k, \pm l)}\rangle)$. 
	In particular, for a given $N_S$, we  track the gap between the single shot quantum Chernoff bounds for the state $|\psi_r^{(\pm k, \pm l)}\rangle$ and the corresponding coherent state with the same signal strength. Mathematically, we are interested in the quantity, which we call  quantum advantage in quantum illumination, given by 
	\begin{eqnarray}
	\Delta(|\psi_r^{(\pm k, \pm l)}\rangle) =   \mathcal{Q}_c - \mathcal{Q}(|\psi_r^{(\pm k, \pm l)}\rangle), 
	\label{eq:qadv}
	\end{eqnarray}
	where $\mathcal{Q}_c$ can be computed  using the same signal strength, $N_S$, as obtained from Eq. \eqref{eq:ns1}.
	If $\Delta(|\psi_r^{(\pm k, \pm l)}\rangle) > 0$,  we report a quantum advantage in QI, while $\Delta(|\psi_r^{(\pm k, \pm l)}\rangle) \leq 0$ implies that  the classical protocol outperforms or performs in a similar fashion as the QI scheme, for the given quantum probe $\rho_{IS}$.\\

	There exist bounds in several quantum information protocols, like dense coding, teleportation, and quantum cryptography, for which  quantum advantage is system-independent and several attempts have been made to understand the origin of quantum advantage in terms of various quantum correlations \cite{MoreThanEnt6,MoreThanEnt}.  However, to date, no conclusive reasoning behind the underpinnings of quantum advantage in the case of the quantum illumination scheme is obtained. Notice that the enhancements, in this case, depend on the structure of states and are achieved by comparing the quantum protocol with the optimal classical ones.
It is plausible that the advantage is obtained due to the quantum correlation between the idler and signal mode which is not  straightforward to compute, and  hence the inherent quantumness in states responsible for quantum advantage in this process is worth investigating.\\
\textcolor{black}{ In this work, we study the performance of non-Gaussian states in the quantum illumination protocol, where the signal strength $N_S$ for both the non-Gaussian and the classical state are taken to be the same. It is known that for the same signal strength, the TMSV state provides a lower error probability compared to the non-Gaussian PA (PS) states (see  \cite{NGIllu,TMSVbest}). Now given the fact that the non-Gaussian states cannot outperform the TMSV state of the same signal strength, it is natural to ask whether the non-Gaussian states give any quantum advantage at all. The comparison with the coherent state of the same signal strength answers this question. We find that although  the non-Gaussian probes cannot provide an advantage over the Gaussian TMSV state, they can outperform the coherent state probe having the same signal strength, thereby guaranteeing quantum advantage.
Moreover, considering the signal strength as the guiding parameter,  we wish to study the hierarchy among the non-Gaussian states by comparing different non-Gaussian states, i.e., photon-added and -subtracted states with the classical ones having the same signal strength in the illumination protocol.
}\\

 \subsection{Advantage in the detection of additional targets: The operational interpretation of quantum advantage}
\label{sec:app4}
	\begin{figure}
		\centering
		\includegraphics[width=\linewidth]{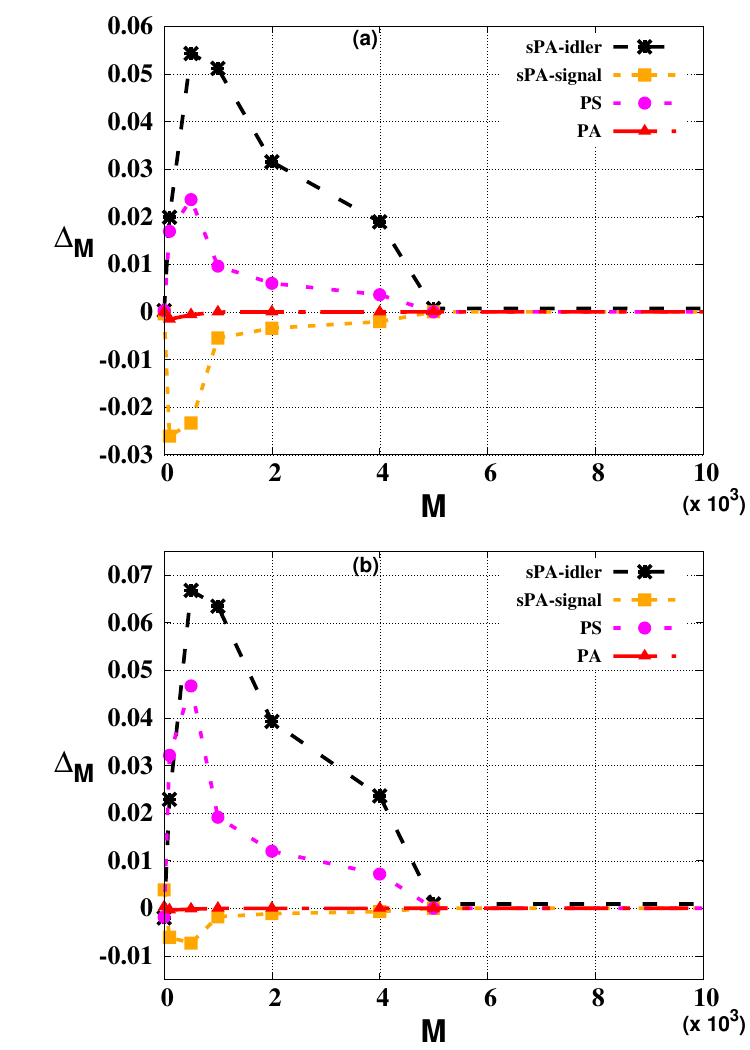}
		\caption{(Color Online.) \textcolor{black}{Quantum advantage in the illumination protocol, \(\Delta_M\) defined in Eq. (\ref{eq:deltaM}), (vertical axis) vs. the number of probe states \(M\) (horizontal axis) for non-Gaussian states created by addition (subtraction) of (a) one photon, and (b) two photons.  All other specifications are the same as in Fig. \ref{cbound_double} for $x = 0.2 ~ (\approx 4dB)$.  Every point shown in the x-axis actually represents a scaling of $10^3$. Both axes are dimensionless.}	
		}
		\label{fig:delta_m}
	\end{figure}
\begin{table*}[htbp]
			\caption{Number of additional targets detected by $M$ copies of the non-Gaussian PS state over the classical probe.} 
			\label{tab:M_additional}
			\centering
		\begin{tabular}{|r|r|r|r|}
\hline
\multicolumn{1}{|c|}{$M$} & \multicolumn{1}{c|}{Coherent state} & \multicolumn{1}{c|}{PS state} & \multicolumn{1}{c|}{No. of extra targets detected} \\ \hline
$1$                       & $59$                                & $63$                          & $4$                                                \\ \hline
$100$                     & $5130$                              & $5468$                        & $338$                                              \\ \hline
$500$                     & $15457$                             & $15928$                       & $471$                                              \\ \hline
$1000$                    & $18968$                             & $19160$                       & $192$                                              \\ \hline
$5000$                    & $19995$                             & $19999$                       & $4$                                                \\ \hline
$\geq 10000$              & $20000$                             & $20000$                       & $0$                                          \\ \hline
\end{tabular}
\end{table*}

	\textcolor{black}{ Let us now go beyond the single copy level, i.e., 
	for $M$ copies of the state, the quantum advantage reads as
	\textcolor{black}{
	\begin{equation}
	    \Delta_M = (\mathcal{Q}_{cM} - \mathcal{Q}_M).
	    \label{eq:deltaM}
	\end{equation}}
Note that it is  a $M$-copy version of \(\Delta\) defined in Eq. (\ref{eq:qadv})}

\textcolor{black}{Let us explain how $\Delta_M$ quantifies the number of extra targets detected by the quantum probe over the classical probe. 
Suppose that the Chernoff bound offered by a coherent state is $\mathcal{Q}_{c} = 0.49852$ and that obtained by using a single copy of  a photon-subtracted state having the same signal strength is $\mathcal{Q}_{PS} = 0.498288$ for target reflectivity $\kappa = 0.01$, and background thermal noise $N_{B} = 1.0$. As a result, the quantum advantage for a single copy of the states is $\Delta = 0.000232$. Let us consider that there are $20000$ targets. The number of additional targets detected by the quantum probe for $M$ copies can be calculated and is given by \textcolor{black}{$((\mathcal{Q}_{cM} - \mathcal{Q}_M) = \Delta_M)\times20000$} (in Eq. \eqref{eq:deltaM}). By varying $M$, we get the statistics given in Table. \ref{tab:M_additional}, \textcolor{black}{where we enumerate the number of targets detected by both the coherent state and the PS state along with the number of additional targets detected by the quantum probe.}}
\textcolor{black}{It may be argued that the advantage evinced through the quantity $\Delta_M$ may as well be represented by the variation of the Chernoff bound against $M$. However, Eq. \eqref{eq:deltaM}, translates the probabilities to the actual number of additional targets that can be detected by the quantum protocol and is thus another heuristic approach towards illustrating the quantum advantage.}

It is observed that, for probes with single photon addition (subtraction) that provide a quantum advantage, the quantum advantage initially increases with $M$ (up to $M \leq 500$), then decreases, and finally vanishes as $M$ becomes very large, as shown in Fig. \ref{fig:delta_m}. For the PA and sPA-signal states, $\Delta < 0$ for $M = 1$. Such states with multiple copies perform even worse as the number of classical probes increases. Finally, for all states, the quantum advantage vanishes in the limit of an infinite number of copies which is expected, since, at that limit, any protocol would eventually be successful. \textcolor{black}{The fact that the number of additional targets detected decreases as $M$ becomes sufficiently large proves that illumination can provide quantum advantage in the limit of a smaller number of probes, e.g., for $20000$ targets, $M \approx 500$ provides a concrete quantum advantage in terms of detecting the extra targets. This manifests another benefit of using quantum states since we need not deploy a large number of states to obtain significant quantum advantage. Moreover, in the case of addition (subtraction) of multiple photons, e.g., $2$ photons as shown in Fig. \ref{fig:delta_m} (b), the trend of $\Delta_M$ is similar to Fig. \ref{fig:delta_m} (a), but with a higher quantum advantage.}


\textcolor{black}{\textbf{Remark.} We note that the Gaussian TMSV state is the optimal resource for the illumination protocol \cite{TMSVbest} in the limit of very low target reflectivity \cite{Lloyd, GIllu}.  As the target reflectivity, $\kappa$, increases, the coherent state probe can easily outperform the Gaussian TMSV state, and hence, at moderate to high target reflectivity, it is sufficient to use a coherent state of high signal strength as the probe. Therefore, non-Gaussian states can never outperform the Gaussian ones in the illumination setup. However, we shall evaluate the conditions under which non-Gaussian states can perform better than their Gaussian counterparts depending on the efficiencies with which photons can be added or subtracted from the TMSV state.}


	

	
	
	
	\subsection{Classification of non-Gaussian states according to their performance in quantum illumination}
	
\begin{figure}
		\centering
		\includegraphics[width=\linewidth]{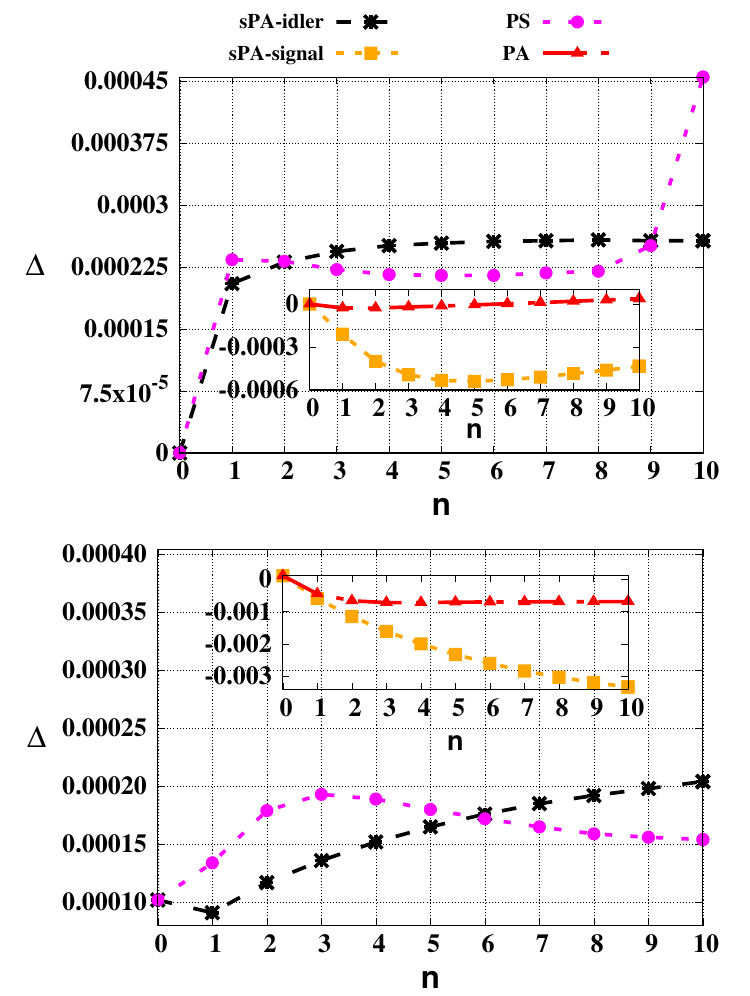}
		\caption{(Color Online.) Quantum advantage  in the illumination protocol, \(\Delta\) defined in Eq. (\ref{eq:qadv}), (vertical axis) vs. \(n\) (horizontal axis).  All other specifications are the same as in Fig. \ref{cbound_double}. Both axes are dimensionless. 	
		}
		\label{cbound_double_Ns}
	\end{figure}

	We are now going to classify the non-Gaussian states according to this figure of merit, \(\Delta\).  We observe that \(\Delta\) is positive for the state with
	photons subtracted from both the modes, thereby establishing 
	quantum advantage in QI 
	while the states with photons added in the idler mode (i.e.,  photon-subtracted from the signal mode) also prove to be a good resource for illumination according to \(\Delta\) (see Fig. \ref{cbound_double_Ns}). 

	Our results indicate that in order to obtain quantum advantage from non-Gaussian operations, it is favorable to perform photon subtraction, either from the signal mode or from both modes. Furthermore, photon-subtraction being easier experimentally, \cite{SubEasy} adds to the importance of our result.

	As it is evident from Fig. \ref{cbound_double_Ns},  all non-Gaussian states, having  $N_S$ equal to the classical signal, do not perform equally well. 
	In particular, 
	the non-Gaussian states with photons added only in the signal  as well as in both the modes, cannot surpass the classical limit which is  clearly visible from the negative values of $\Delta$.\\
	\noindent \textbf{Note $\mathbf{1}$.} The same hierarchies will  be maintained for a higher number of copies of the probe states used for QI which directly follows from Eq. \eqref{eq:deltaM}.\\
	\noindent \textbf{Note $\mathbf{2}$.} Such a definition of quantum advantage   also has limitations. In particular, it ignores the probabilistic nature of the   photon addition (subtraction) operation  in (from) the TMSV state. \\
	\noindent \textbf{Note $\mathbf{3}$.} In presence of low reflectivity, TMSV   is found to be better than the corresponding photon-added and -subtracted states, having the same signal strength \cite{TMSVbest}.  To match the signal strength for both states,  the squeezing parameter of the TMSV turns out to be higher.  This feature is not unique to QI, but also happens in the case of entanglement as well, where for a given average number of photons, the TMSV state gives the maximal amount of entanglement \cite{NGIllu} over any photon-added (- subtracted) state.\\
	\textcolor{black}{In Sec. $4$ of the Supplementary material, we analyze how different non-Gaussian states perform in the illumination protocol when the reflectivity of the target is not correctly estimated.}
	
	

	\section{Chernoff bound vs. quantum advantage in illumination with noisy non-Gaussian probes}
	\label{sec:noisy}
	
	In this section,  we consider the performance of entangled states in presence of different kinds of noise in resource states, as discussed in Sec. \ref{subsec:noise}.  Till now,   noise is considered as a strong  thermal background, present around the target. 
	Here, we focus on  a situation when
	the two-mode photon-added and  -subtracted entangled states,  admixed with i) local noise, ii) generated via faulty twin beam generator, iii) having imperfect photon addition or subtraction,  are used as a probe. We investigate the behavior of CB as well as the quantum advantage in QI, defined in Eq. (\ref{eq:qadv}), for these noisy probes.   
	In spite of the worsening performance in QI in presence of noise,  we report  several advantages exclusive to non-Gaussianity such as robustness against noise in QI and activation of quantum advantage in CB. 
	
	

	
	
	\subsection{Chernoff bound against local noise in non-Gaussian states}
	\label{sec:localn1}
	
\begin{figure}
		\centering
		\includegraphics[width=\linewidth]{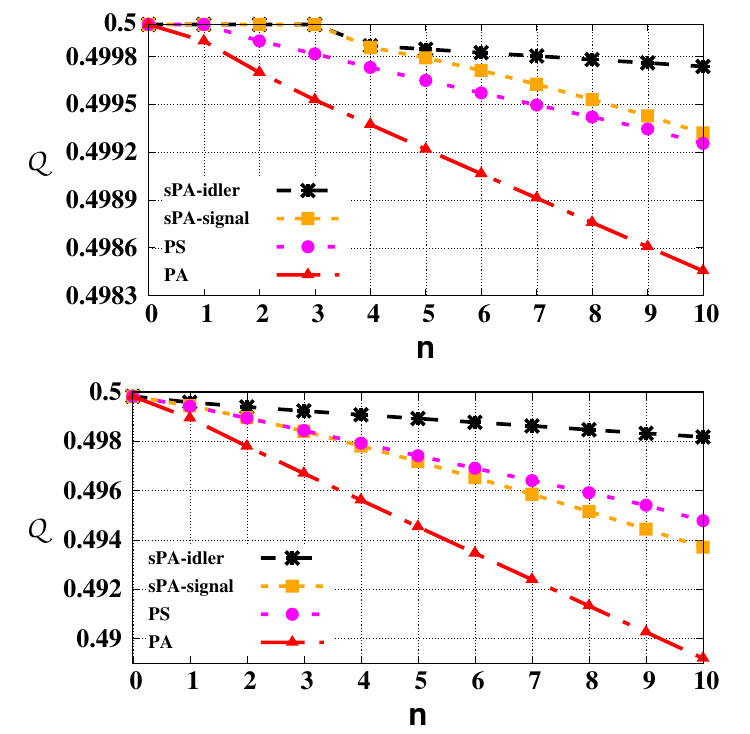}
		\caption{(Color Online.) Robustness and activation of CB against noise. 
			\(\mathcal{Q}\) (vertical axis) with respect to \(n\) (horizontal axis)  when probes are mixed with local Gaussian noise, given in Eq. (\ref{noisy_state}). The upper panel represents states with $p = 0.9$  (which shows activation of Chernoff bound in case of addition of photons in the idler or signal mode) while the lower panel corresponds to $p = 0.3$. Here \textcolor{black}{$x = 0.2 ~ (\approx 4dB)$}. 
			All other specifications are the same as in Fig. \ref{cbound_double}. 
			Both axes are dimensionless.	}
		\label{gaussian_robust}
	\end{figure}

	Let us first consider the local noise models given in  Eqs. \eqref{noisy_state} and \eqref{gaunoise}.
	The robustness against noise is characterized by the maximum value of the mixing probability, $p = p^{*}$, in Eq. (\ref{noisy_state}) below which the CB, $\mathcal{Q}$, is lower than \(0.5\) (up to numerical accuracy of the order of \(10^{-4}\)). \textcolor{black}{An increasing value of $p^{*}$ indicates \textit{enhanced robustness to noise}, (see Fig. \ref{gaussian_robust}).}  Before moving to the photon-added and -subtracted states, let us first notice that for the TMSV state, \(p^{*}\) turns out to be \(0.5\) when \textcolor{black}{ $x = 0.2 ~ (\approx 4dB)$} while 
	it reduces to $0.4$ for low squeezing strengths, \textcolor{black}{$x = 0.05 ~ (\approx 2dB)$}. 
	
	\subsubsection{Enhanced robustness against noise}
	
	 We illustrate our results for  two exemplary squeezing parameters, $x$,  although the results remain qualitatively similar for other squeezing parameters:
	
	\begin{enumerate}
		\item For \textcolor{black}{$x = 0.2 ~ (\approx 4dB)$}, when photons are added or subtracted in the signal or in the idler, $p^{*}$  increases to $0.7$ from $0.5$, indicating enhanced robustness. When more photons are added (subtracted) in a single mode, and when two mode operations have been employed,  $p^{*}$ increases further.

		\item Such robustness decreases with the decrease of the squeezing parameter of the original TMSV state used for non-Gaussian operations. For example,  when \textcolor{black}{$x = 0.05 ~ (\approx 2dB)$},  the tolerance against noise increases for single mode operations,  going from  $p^{*} = 0.4$ in the TMSV Gaussian state, to $p^{*} = 0.6$ for the non-Gaussian states, i.e., \(p^{*}_{nG} > p^*_{TMSV}\).   The trend is more prominent when photons are added (subtracted) in both modes. 
	\end{enumerate}

	\noindent	\textbf{Note.}  Since $\mu_n$ and $\nu_m$ represent the \textcolor{black}{ probabilities of having $n$ and $m$ photons in a mode respectively},  if they increase, so does the average number of noisy photons, and thus the performance of CB becomes poor. In the case of the Gaussian noise model, the mean number of noisy photons increases with an increase of $\sigma_1$ and $\sigma_2$ in Eq. \eqref{gaunoise}, thereby providing higher error probability with the increase of noise.
	
	\subsubsection{Activation of Chernoff bound via non-Gaussianity}
	\label{sec:acti1}
	
	Another interesting feature of non-Gaussianity is that, for  noise strengths that undermine the performance of the TMSV state such that it is not better than blindly guessing the target, photon-added and -subtracted states  give a reasonable CB (which is much below \(0.5\)). It means that the non-Gaussianity can help to counter the adverse effect of noise and lead to activation of \(\mathcal{Q}\). For example, with \textcolor{black}{\(x =0.05 ~ (\approx 2dB) \)},  for local noise, the TMSV state cannot outperform the blind guess probability, $p = 0.5$, while the states with photons added in the idler mode, show  activation for \(n \geq 3\) (see Fig. \ref{gaussian_robust} (upper panel)) and with \textcolor{black}{\(x = 0.2 ~ (\approx 4dB)\)}, \(n \geq 2\) is enough.  

	\subsection{Quantum advantage with  noisy non-Gaussian probes}
	
	In presence of  Gaussian noise, we find that the quantum advantage in terms of positive \(\Delta\) can  not be achieved using the photon-added and  -subtracted states if the noiseless coherent and noisy non-Gaussian states possess the same signal strength. Specifically, in this scenario, \(\Delta (\rho) \leq 0\).
	
	Instead of comparing the performance of noisy non-Gaussian states with the optimal classical scheme by coherent states, 
	let us  consider a scenario where the signal transmission line itself is affected by noise and hence any state passing through it suffers from same amount of noise, i.e., noisy channel affects both coherent and photon-added (subtracted) states.
	When local Gaussian noise 
	acts on it, a coherent state, transforms in the following way:
	\begin{equation}
	\rho = (1 - p) |\omega \rangle \langle \omega| + p \sum_{n = 0}^\infty \mu_n |n \rangle \langle n|
	\label{coherent_noise}
	\end{equation}
	where $|\omega \rangle = e^{-\frac{1}{2}|\omega|^2 } \sum_{n = 0}^\infty \frac{\omega^n}{\sqrt{n!}}|n \rangle \langle n|$ is the coherent state and $\mu_n$ is given in Eq. \eqref{gaunoise}.
	The signal strength 
	is then given by
	\begin{eqnarray}
	N_S = (1-p)|\omega|^2 + p \sum_{n = 0}^\infty n \mu_n.
	\label{eq:nscoherent}
	\end{eqnarray}
	When the target is present, 
	the state $\rho_1$ (as in Eq. \eqref{rho_1_exp}) reads 
	\begin{eqnarray}
	&& \hspace{-1em} \rho_1 = e^{-|\omega|^2}\sum_{n,n' = 0}^{\infty} \sum_{m = 0}^\infty \sum_{r = 0}^{n }\sum_{s,s' = 0}^{m}
	f\left(\{n\}, \{r\},\kappa,N_B\right) \times
	\nonumber  \\  
	&& \hspace{-1em}
	~~~~~~~~~~~~~~~~\ket{ n + m - r - s}\bra{ n' + m  - r - s}_{S}, ~~~~
	\label{rho1_coh}
	\end{eqnarray} 
	where $f\left(\{n\}, \{r\},\kappa,N_B\right)$ has the same form as Eq. \eqref{prefactor} with $k = l = 0$ and $c_n = \frac{\omega^n}{\sqrt{n!}}$. 
	Since the coherent state constitutes a single-mode probe, in the absence of the target, the signal is lost and all that the detector receives is the background thermal noise i.e., $\rho_0 = \rho_T$. We evaluate the Chernoff bound for the noisy coherent state using these expressions.
	
	\begin{figure}
		\centering
		\includegraphics[width=\linewidth]{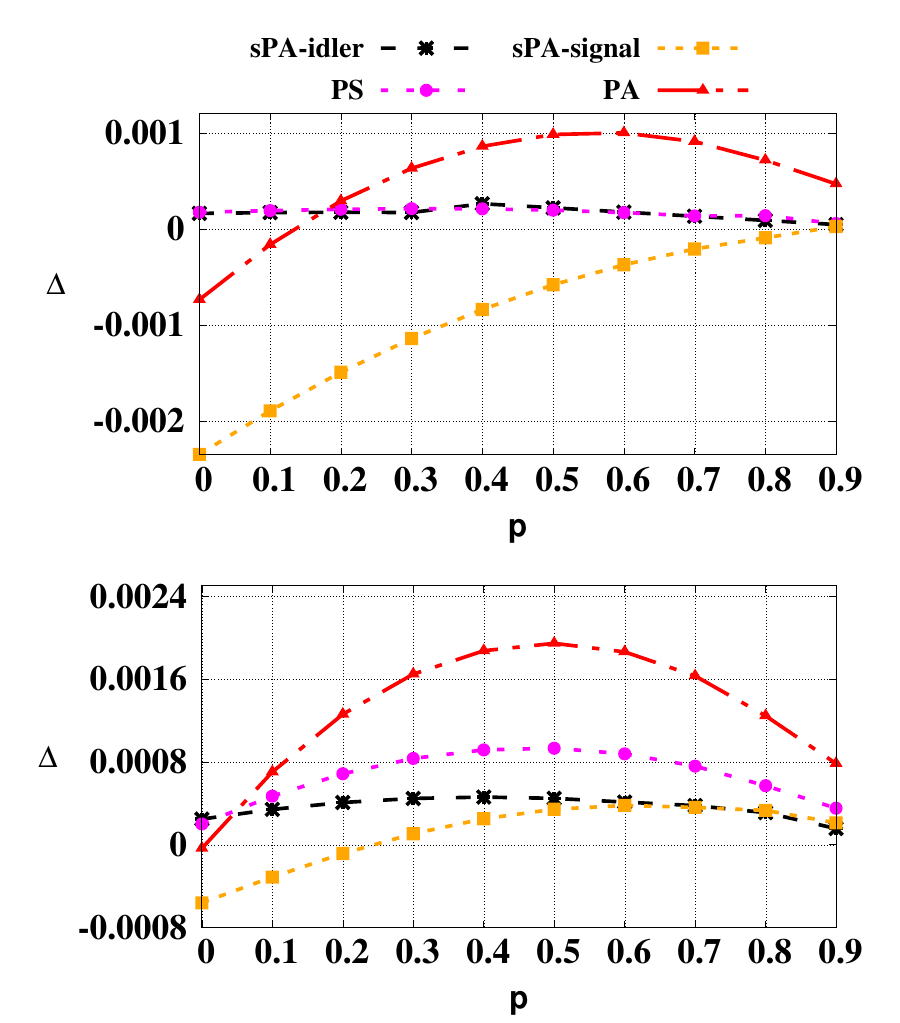}
		\caption{(Color Online.)
			Robustness of quantum advantage, \(\Delta\),  (ordinate) with the variation of noise strength, \(p\) (abscissa). In Gaussian noise,  $\sigma_1 = \sigma_2 = 1$.   In all cases, the number of photons added or subtracted is set to be $5$.  We fix \textcolor{black}{$x = 0.2 ~ (\approx 4dB)$} in the lower panel and \textcolor{black}{$x = 0.05 ~ (\approx 2dB)$} in the upper panel. All other notations are the same as in Fig. \ref{cbound_double_Ns}. Both axes are dimensionless.}
		\label{gau_robust_fig_Ns}
	\end{figure}
	

	We compare now the noisy non-Gaussian states with the corresponding noisy coherent state, i.e.,  noise affects both non-Gaussian and coherent states in a similar fashion, so that
	the signal strength gets modified in exactly the same way as in Eq. \eqref{eq:nscoherent}. For the comparison under the same signal strength, the choice of the intensity of the coherent light can be chosen using Eqs. \eqref{eq:nscoherent} and \eqref{eq:ns1} by
	\begin{eqnarray}
	|\omega|^2 = \sum_{n =\frac{k - (\pm k)}{2}}^\infty (n\pm k) ~|c_n^{(\pm k, \pm l)}|^2
	\label{eq:coh_intensity}
	\end{eqnarray}
	which is nothing but the signal strength of the entangled  mixed state given in Eq. \eqref{noisy_state}. \textcolor{black}{For the comparison of illuminating power, we choose a coherent state of intensity given by Eq. \eqref{eq:coh_intensity}.}
	It is evident from Fig. \ref{gau_robust_fig_Ns} that  $\Delta >0 $ for all non-Gaussian states for a certain range of \(p\). This is due to the fact that with the increase of noise, the classical resource is much more affected by the noise and hence its performance degrades drastically.  However, at low levels of noise, the state with photons added in the signal mode performs poorly as compared to the classical probe (as shown in Fig.  \ref{gau_robust_fig_Ns} where  \(\Delta\) of this state  becomes positive when  $p \geq 0.3$, from the negative value in presence of  low noise). With the increase of the noise strength, all the  non-Gaussian resources  perform much better than that of the classical probe until the noise probability becomes too large ($\geq 0.5$) and $\Delta$ starts to decrease again, although remaining positive. Thus we demonstrate that under the destructive effect of noise, the non-Gaussian probes are  more robust than  the noisy classical probes having the same signal strength, thereby exhibiting quantum advantage. In this scenario, the state with photons added in both modes constitutes the most robust resource. The results remain qualitatively similar even when the squeezing strength, \(x\) is moderately small. \\
	
	\textbf{Note.} Regarding the understanding of enhancement features of non-Gaussian states, non-Gaussianity measures can be employed to answer the increment obtained in non-Gaussian states via photon addition and subtraction \cite{Cerfaddsub,tamo2016, SappyTamoGBell,ratulggm}. However, it fails to conclusively identify the reason behind the fact that subtraction is better than addition or  vice versa. In the case of quantum illumination process which depends on systems critically, since the advantage can only be ensured by comparing the quantum protocol with the classical one, it is hard to pinpoint a single quantifier of states which can capture the advantages reported here. However, as mentioned before, the results strongly indicate that the quantumness present in the process via the correlations in the signal-idler pair is possibly responsible for any quantum advantage.

	\subsection{Effects of faulty twin beam generator on quantum efficiency}

	As described in  Sec.\;\ref{subsec:faulty}, we  consider the effects of a faulty twin beam generator, which produces TMSV with \(x'< x\) (where \(x\)  is the promised squeezing strength), on the CB. Notice that a similar investigation can be carried out with  \(x' > x\) which we are not considering since states with higher squeezing are typically hard to prepare compared to those with a lower one.   The Chernoff bound for the non-Gaussian states is calculated for squeezing parameter $x$, thereby yielding $\alpha_x$ as its optimal parameter. However, due to the faulty twin beam generator, we obtain the TMSV state having squeezing $x'$ which we are unaware of. Hence during the computation of CB, we apply optimal  $\alpha_{x}$ in Eq. (\ref{eq:CBdef}), instead of \(\alpha_{x'}\), thereby leading to a higher error probability compared to that obtained via \(x\). We refer it as \(\tilde{\mathcal{Q}}\).  Notice  here that    CB increases, with the decrease of  $x$  of TMSV, thereby  increasing the probability of error.

	\begin{figure}[ht]
		\centering
		\includegraphics[width=\linewidth]{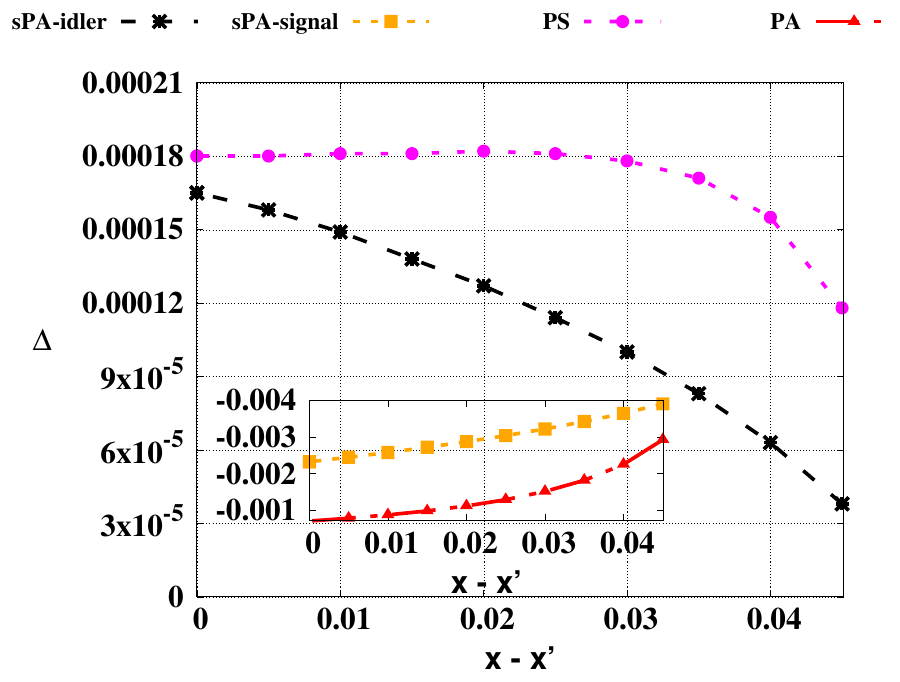}
		\caption{(Color Online.) 
			Role of faulty twin beam generator on the performance of illumination. Quantum advantage, \(\Delta\), (vertical axis) against $x - x'$ (horizontal axis).  All other specifications are the same as in Fig. \ref{cbound_double_Ns}. Both axes are dimensionless.  }
		\label{bad_squeeze_Ns}
	\end{figure}
	

	\textit{Quantum advantage due to faulty twin beam generator.}
	For this analysis, let us fix \textcolor{black}{$x = 0.05 ~ (\approx 2 dB)$} and choose various values of $x' < x$.  Here we consider coherent states with $N_S$ corresponding to $x$, since $x'$ is an artifact of the faulty device and beyond the scope of the experimentalist. 
	As illustrated in Fig. \ref{bad_squeeze_Ns}, in presence of a low faulty twin-beam generator, i.e. when $x - x' < 0.03$,  the positivity of \(\Delta\) guarantees the quantum advantage by the photon-subtracted  and the photon-added states in the idler mode while the addition of photons in both the modes and in the signal modes fail to show any advantage. The beneficial nature of non-Gaussian states disappears with the slight increase of \(x-x'\). Notice also that the results remain similar for other values of the squeezing parameter.   
	
	
	\textcolor{black}{Note that quantum illumination eventually boils down to a state discrimination problem between two final states
- one obtained when the target is present and the other in the case of the absence of the target. State discrimination protocols are based on the knowledge of which states need to be distinguished. Therefore, the final state discrimination efficiency, quantified by the Chernoff bound, is calculated based on the prior knowledge of the probe states. In particular, we can consider that two POVM operators distinguish the two states with some error, and this error is bounded by $\mathcal{Q}$. So for a given state in mind, the experimenter sets the two optimal POVMs that are used for the discrimination scheme. The error probability, which is calculated theoretically depending on the optimal choice of $\alpha$, should match with the practical scenario if there is no error in preparation. However, when there is a faulty twin beam generator, unbeknownst to the experimenter, the optimal POVM cannot be changed and instead, the previous choice of POVM is used for state discrimination, which is not optimal. This translates to evaluating the Chernoff bound with a wrong exponent that corresponds to the expected state.} We wish to see how a change in the squeezing strength from the desired value affects the illumination protocol. We recall that the Chernoff bound for the TMSV state depends on the signal strength as $\mathcal{Q} = \exp{(-\kappa N_S/N_B)}$ \cite{GIllu} and thus on the squeezing parameter (since $N_S = 2\sinh^2 x$). Therefore, if the squeezing strength is different from what is required, the error probability will be much worse than expected, \textcolor{black}{since the discrimination will be performed in terms of the POVMs for the expected squeezing amplitude}. Given a probe state of a fixed squeezing strength, the experimenter has an idea about the error probability that can occur in the protocol and thus designs the setup accordingly. A change in the squeezing strength due to the faulty twin-beam generator would then affect the successful detection probability. It is this change in the error probability that we are concerned with. Thus treating the unexpected change in the squeezing strength as a fault (noise) in the protocol, we study its effects on the eventual outcome. \\
 Notice, moreover, that errors in the illumination protocol can not only arise from state imperfections but also due to inefficiencies in the detection process. The study of quantum illumination from the perspective of detectors, however, is an entirely different avenue of research. Our discussion in this manuscript is focused on the aspect of state preparation and errors during signal propagation, which are also important components of the illumination protocol. Several works have focused on the detection schemes in illumination alone \cite{Karsa_IEEE_2020,Jo_PRR_2021} and such a description is beyond the scope of our present manuscript. \textcolor{black}{In Sec. $5$ of the Supplementary material, we describe the effect of another kind of noise, that deals with imperfect photon addition or subtraction.}



	\section{Probabilistic improvements via non-Gaussianity}
	\label{sec:advantage_ng}
	
	\textcolor{black}{In the first four sections, we have demonstrated whether a single copy of the \textcolor{black}{non-} Gaussian resource can outperform a single copy of the coherent state probe having the same mean photon number, irrespective of the success probability of creating the \textcolor{black}{non-} Gaussian state, thereby arguing that a more just guiding parameter for comparing different probes is their respective signal strengths.  This treatment was motivated by that in Ref. \cite{NGIllu}, where the performance of Gaussian and non-Gaussian states are compared at the same squeezing strength, irrespective of their probability of generation. }
 
 In this section, we employ another procedure to compare the efficiency of the QI method by Gaussian and non-Gaussian states. In particular, we consider the indefiniteness involved in the photon addition and subtraction operations in which a fewer number of non-Gaussian states are distilled from a higher number of TMSV states. \textcolor{black}{For a given distillation rate, we intend to compare the performances of the base Gaussian TMSV state and the distilled non-Gaussian states.} Importantly, this is different from the imperfection in the photon addition (subtraction) mechanism, as we will discuss shortly \textcolor{black}{(see Supplementary Material Sec. $5$ for the illumination protocol affected by imperfect non-Gaussian operations)}. Our comparison scheme in the first subsection takes into account the indeterminacy in the photon addition and subtraction process, both in the presence and absence of noise.

	\subsection{Inefficient non-Gaussian apparatus}
	
	Photon addition and subtraction are not deterministic processes (cf. \cite{Cerfaddsub} and references therein). Their successful implementation is denoted by some kind of clicks that indicate the success of the process. It has been established that \cite{add-sub-prob,Cerfaddsub,tamo2016,SappyTamoGBell,ratulggm} photonic operations to create non-Gaussian states are interpreted as a distillation process where one post selects the states that had a positive click, i.e., states for which the relevant photonic operation was successful. Therefore, starting from $M$-copies, one ends up with a lesser number ($M'$-copies) of more entangled states. States for which the photonic operations fail are rejected.  The success probability is simply the ratio $\eta = M'/M$. In our work, since QI is explicitly dependent on the number of copies used to probe the target, we compare the performance of the $M$-copies of the base Gaussian states with $M'$-copies of the distilled non-Gaussian states and check which one is better with respect to the success probability $\eta$.

\begin{figure}
		\centering
		\includegraphics[width=\linewidth]{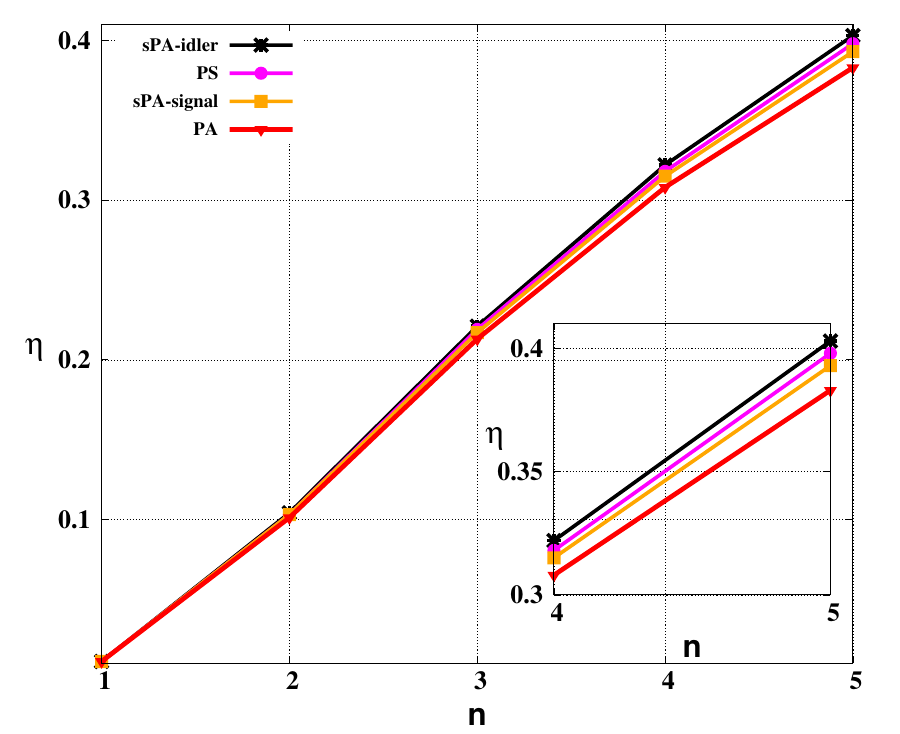}
		\caption{(Color Online.)  \textcolor{black}{Minimum operational efficiency ($\eta$) required (ordinate) to add (subtract) a single photon such that a final $n$ photon-added (-subtracted) state  (abscissa) can outperform a higher number of Gaussian states with \textcolor{black}{$x = 0.2 ~ (\approx 4dB)$}, given in Eq. (\ref{eq:response_eq_1}). The probes are pure non-Gaussian states. The inset shows the hierarchy in the variation of $\eta$ when we consider final states with a high number of added (subtracted) photons All other symbols used here are the same as in Fig. \ref{cbound_double}. Both axes are dimensionless.  }}
		\label{pN}
	\end{figure}

\textcolor{black}{We now illustrate that even if the apparatus is not 100\% efficient in preparing photon-added and -subtracted states, non-Gaussianity can offer an added advantage in the illumination protocol, as compared to the TMSV states' performance. The main idea is to find the minimum probability with which a given non-Gaussian state needs to be prepared such that it can overcome the error bound due to the TMSV probe. To this end, let us consider that we have $M$ copies of the TMSV state. The Chernoff bound for the same is $\mathcal{Q}^M_{TMSV}$. Let us also consider that the probability of adding (subtracting) a single photon to a state (Gaussian or non-Gaussian) is $\eta$. Then the probability of successfully obtaining a single photon-added (subtracted) state is $\eta$, that of obtaining a two photon-added (subtracted) state is $\eta^2$, and so on, with the probability of successfully generating an $n$ photon-added (subtracted) state being $\eta^n$. \textcolor{black}{This is a reasonable assumption, e.g., at $x = 0.2~ (\text{or}~ r = 0.48)$, the probability of successfully subtracting $n$ photons scales approximately as the $n$-th power of subtracting one photon (see Table $1$ in Ref. \cite{referee_suggest})}. Therefore, we can calculate the minimum preparation probability $\eta$ required for $\eta^n M$ copies of an $n$ photon-added (subtracted) state such that it can furnish a lower error probability than $M$ copies of a TMSV state, i.e.,
 \begin{equation}
     \mathcal{Q}^M_{TMSV} = \mathcal{Q}^{\eta^{n}M}_{non-Gaussian}.
     \label{eq:response_eq_1}
 \end{equation}
\textcolor{black}{Therefore, we calculate the required probability, $\eta$, of adding (subtracting) one photon to generate the non-Gaussian probe so that Eq. \eqref{eq:response_eq_1} is satisfied, i.e., the Chernoff bound for $M$ copies of the TMSV is equal to that of $\eta M$ copies of the non-Gaussian state. This denotes the minimum probability of success required in the generation protocol such that non-Gaussian states can perform better than their parent Gaussian state. Note that we do not assume the probability of success of the non-Gaussian resource and instead, calculate the required success probability of generation such that the non-Gaussian state can offer a lower Chernoff bound than $\mathcal{Q}^{M}_{TMSV}$. Our analysis sheds light on the experimental accuracy required to create a lesser number of non-Gaussian states that can perform better than a larger number of Gaussian resources.} Our only assumption in this regard is that the probability of de-Gaussification is the same in both modes of the parent state. We demonstrate our results in Fig. \ref{pN} which shows the variation of $\eta$ with the number of photons that the final non-Gaussian state must have added (subtracted). } When the photonic operation fails,  we are left with a junk state which is useless for any information processing purpose.
	
	Our results indicate that the minimum operational efficiency required by the non-Gaussian apparatus to add (subtract) a single photon, must increase with an increase in the total number of photons to be added (subtracted). This is also consistent with experiments since it is more difficult to introduce higher non-Gaussianity in a state.  For final states with a small number of total photons added or subtracted, the efficiency required is nearly the same and also quite low (for example, to prepare states with $2$ added (subtracted) photons, $\eta \leq 0.1$.)
	It turns out that the addition of photons to the idler mode demands the most efficiency while
	the subtraction of photons from both modes succeeds the former type of states, followed by the addition of photons in the signal mode. Thus instead of subtracting photons from both modes, we can deal with the addition of the same to the signal mode to obtain a low CB. 
	
	\textcolor{black}{\textbf{Note}. Fig. \ref{pN} may give an impression that the addition (subtraction) of a large number ($n \geq 5$) of photons is experimentally feasible. However such a high value of $n$ is just a  theoretical construct to show how the Chernoff bound and the quantum advantage scales with the number of added (subtracted) photons although, with current experimental status, such a number of photon addition (subtraction) is not possible in laboratories. 
Furthermore, the success probability, $\eta$, is the probability required for single photon addition (subtraction) such that the final states prepared may beat the TMSV state-based protocol at the same squeezing strength which is also higher than the present experimentally possible probabilities (see Table. 1. in \cite{referee_suggest}). It means that the advantage discussed here may not yet be achieved in laboratories although 
we firmly believe that  states with a small number of added (subtracted) photons with the required success probability will be possible to obtain in near future.
}

	\textcolor{black}{Let us now justify our assumption that when the photonic operations fail, the ``junk'' states which are left must be rejected. Since coherent states are a resource in the illumination protocol, supplementing the ``junk'' states (when the photon addition (subtraction) operation fails) with coherent states is costly and increases the resource overhead. Therefore, we consider only probabilistic non-Gaussian illumination in our analysis. On the other hand when non-Gaussian states are prepared in a heralded manner, one obtains TMSV states with probability $1 - \eta$. Therefore, in the event of experimental failure, one can employ such TMSV states for illumination. However, since the Chernoff bound is not a convex function, it is analytically challenging to calculate the error probability when non-Gaussian states are used with probability $\eta$ and the TMSV is used with probability $1 - \eta$. We leave this problem as an open question for further research.} \\

 \section{ Role of correlations on  quantum illumination}
	\label{sec:app6}
	\begin{figure}[h]
		\centering
		\includegraphics[width=\linewidth]{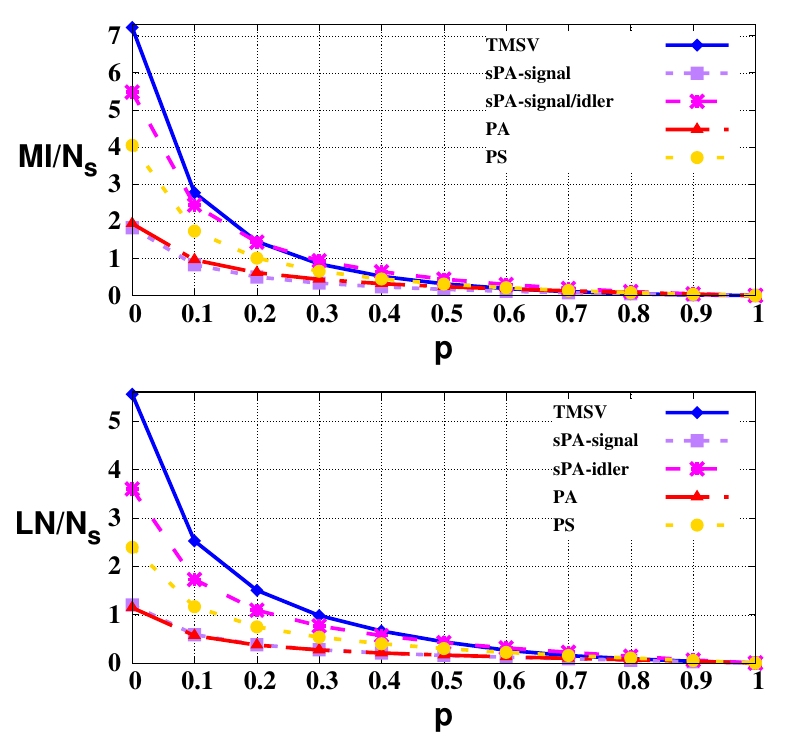}
		\caption{(Color Online.)  Correlations per signal photon (ordinate) against  Gaussian noise strength, \(p\). 
			(Upper panel) Mutual information ($MI$) and (lower panel) logarithmic negativity.  Diamonds correspond to TMSV states and 
			other legends are the same as in Fig. ($2$) of the main text. 
			For all non-Gaussian states, the number of added (subtracted) photons in each mode is $n = 5$.}
		\label{mi_e_Ns}
	\end{figure}
	
\textcolor{black}{We now investigate the role of correlations between the two modes of the probe state on the illumination protocol. The correlations are quantified by the mutual information (MI) \cite{MI1,MI2,MI3} which is defined as $ MI = S(\rho_A) + S(\rho_B) - S(\rho_{AB})$ for a two-mode state $\rho_{AB}$ where $S(\rho)$ denotes the von-Neumann entropy of a quantum state and thus entanglement present per signal photon in the probe states. In our analysis, entanglement for mixed state is quantified via logarithmic negativity \cite{LN1,LN2}, defined as $LN= \log_2 (2N(\rho_{AB})+1)$, where $N(\rho_{AB})$, known as \textit{negativity}, is the sum of the absolute value of the negative eigenvalues of the partial transposed version $\rho_{AB}^{T_A}$ of the two-mode state. Let us carry out the investigation for the noisy probes as specified by Eq. \eqref{noisy_state}. \\
	We find that the Gaussian TMSV state possesses the maximum amount of  correlation, both in terms  of \(MI/N_s\) and $LN/N_s$ as depicted in Fig. \ref{mi_e_Ns}. This may be an indication of such a state being the optimal probe for  illumination purposes \cite{NGIllu}. Interestingly, among the non-Gaussian states, the states with photon-added in the idler and the ones with photons subtracted from both the modes have a higher  correlation with respect to $p$ per signal photon than that of the states having photon addition in signal, and in both the modes. 
	Specifically, non-Gaussian states which cannot furnish any quantum advantage over coherent states having the same $N_s$ lie at the bottom of our figure. Moreover,  as expected, QCs decay with the increase of noise.
	Therefore, a hierarchy in QCs observed in	Fig. \ref{mi_e_Ns} is in good agreement with the quantum advantage obtained for the QI process, thereby connecting the inherent property of the quantum states with the illumination. \\
		Interestingly, unlike most other quantum information protocols, it is not clear whether the entanglement content of the state leads to the quantum advantage in QI. In particular, we know that for QI with the TMSV state, in the limit of large signal strength, the quantum advantage vanishes \cite{TrepsArxiv}, although the entanglement content, $E$,  of the pure state quantified by the von-Neumann entropy of local mode (entanglement entropy), diverges. We now check how the normalized version, $E/N_S$, behaves in the large $N_S$ limit.
	For the TMSV state, $|\psi_r\rangle$, we know
	\begin{eqnarray}
	E(|\psi_r\rangle) &=& \cosh^2 r \log \cosh^2 r - \sinh^2 r \log \sinh^2 r, \nonumber \\
	\label{eq:e/ns}
	\end{eqnarray}
	and since \( N_S(|\psi_r\rangle) = \sinh^2 r\), the entanglement reduces to
	\begin{eqnarray}
	E = (N_S+1) \log (N_S+1) - N_S \log N_S.
	\end{eqnarray}
	Moreover, using Eq. \eqref{eq:e/ns}, we have
	\begin{eqnarray}
	\lim_{N_S \rightarrow \infty} \frac{E}{N_S} = \lim_{r \rightarrow \infty} \coth^2 r \log \cosh^2 r - \log \sinh^2 r = 0. \nonumber \\
	\label{eq:limit}
	\end{eqnarray}
	We know that for a given $N_S$, the TMSV provides the maximal entanglement since its reduced subsystems are thermal. Recall, the thermal states are the ones that yield the maximum entropy for a given temperature (average number of photons), and, therefore, for a fixed signal strength, the entanglement entropy content of the TMSV state is maximal among all other pure states. For any other state with a given value of $N_S$, the entanglement content, $E'$ satisfies
	\begin{eqnarray}
	E' < (N_S+1) \log (N_S+1) - N_S \log N_S = E.
	\end{eqnarray}
	Consequently, from Eq. \eqref{eq:limit}, we can see that in the asymptotic limit,
	\begin{eqnarray}
	\lim_{N_S \rightarrow \infty} \frac{E'}{N_S} = 0.
	\end{eqnarray}
	Physically, the $N_S$ value is highly dependent on the reflectivity, $\kappa$, and the number of copies, $M$ used in QI. For example, in the case of the TMSV, $N_S = 1$ can be considered large for $M =1000$ and $\kappa = 0.1$, while for the same value of $\kappa$ but $M = 1$, $N_S$ has to be taken to a much higher value than unity to be considered as ``large''.  \\
	Nevertheless, we clearly observe that, unlike entanglement, the normalized version of it provides consistent results in the asymptotic case as well, and for large signal strengths, it clearly predicts vanishing quantum advantage for any state considered for QI. When one couples this fact with its accurate predictions of hierarchies of states based  on the scale of quantum advantage, the normalized version of  correlations makes a strong case for themselves on having a deep connection with the quantum advantage obtained in the QI protocol.}

	
	\section{Conclusion and open problems}
	\label{sec:conclusion}
	
\textcolor{black}{Quantum illumination (QI) utilizes shared entanglement between the signal and the idler (an additional mode that is directly sent to the detector) to enhance the detection probability of a target embedded in a thermal background. In addition to important theoretic advantages, several experiments \cite{Lopaeva_PRL_2013,Zhang_PRL_2013,Zhang_PRL_2015} have also been performed both in optical \cite{Balaji_IEEE_2018,England_PRA_2019} and microwave regimes \cite{Barzneh_Science_2020, Assouly_NP_2023} exhibiting quantum advantage in  illumination. }
One of the most recent endeavors in the theoretical front is to use  non-Gaussian states for quantum illumination. 
In our work, the non-Gaussian states are obtained by adding or subtracting photons from the two-mode squeezed vacuum state, an efficient mechanism to generate non-Gaussian states  experimentally.  Our aim was to  categorize non-Gaussian states  based on their performance in quantum illumination which, in turn, reduces to the discrimination of two states.

\textcolor{black}{The fact that non-Gaussian states perform better than their Gaussian counterparts was demonstrated in Ref. \cite{NGIllu} having equal squeezing strength. Specifically, with the same signal strength, non-Gaussian states can never outperform the Gaussian TMSV state. Although this result is counter-intuitive and does not align with other enhancement features reported in the literature induced by non-Gaussianity for the cases of entanglement \cite{Cerfaddsub,tamo2016}, Bell inequality violation \cite{SappyTamoGBell}, etc., we argue that the figure of merit in quantum illumination is qualitatively different. It differs primarily because the Chernoff bound explicitly depends on the signal strength. Therefore, any advantage of non-Gaussianity must be analyzed for a fixed signal strength. Intuitively, this comparison scheme might yield a different result for the following reasons. Firstly, note that non-Gaussian operations almost always increase the average number of photons (signal strength). Consequently, when a performance comparison is made for the same signal strength, the Gaussian counterpart must also have the same signal strength as the non-Gaussian one. It cannot be the Gaussian state from which the non-Gaussian state was obtained as done in Ref. \cite{NGIllu}. Therefore, we believe that our treatment of non-Gaussian quantum illumination, where the signal strength is considered to be the guiding parameter, sheds new light on the theory of remote target detection.}

\textcolor{black}{In our work, we showed that even though all non-Gaussian states can furnish a low enough Chernoff bound, only the photon-subtracted state and the idler-photon-added state can outperform the coherent state having the same signal strength. We believe that since the performance of the coherent state heavily depends on the signal photon number \cite{GIllu}, comparison based on the signal strength is the way to properly access the quantum advantage in the illumination protocol. Thus, we have focused only on the performance of non-Gaussian states and compared their performance with respect to the coherent state probe. The non-Gaussian probes cannot provide an advantage over the Gaussian TMSV state, even though they can easily outperform the coherent state probe having the same signal strength. We also illustrated how converting a large number of copies of the Gaussian probes into a smaller number of non-Gaussian ones can help to obtain a better quantum advantage.}\\

In any experimental implementation, noise is inevitable, and in our work,  the effects of different noisy probe states generated via different imperfections on the illumination procedure are investigated. Considering local noise modeled by Gaussian distributions, we found that, unlike a noiseless scenario, if the signal transmission line equally affects both the non-Gaussian and coherent states having equal signal strength, all non-Gaussian states give a quantum advantage. Specifically, in the presence of certain critical noise values, benefits via non-Gaussian states increase with the increase of noise. 
	In addition, we considered faulty twin beam generators producing two-mode squeezed states having lower squeezing strength than the promised one. In all these situations, photon-subtraction in both the modes and in the signal mode always gives improvements in QI. 

Our investigations also open up the possibility to include more realistic losses in the setup including propagation losses,  divergence, atmospheric attenuation, and other losses.  In such a situation, one can argue that the chosen reflectivity in our study is actually the effective reflectivity of the target seen by the signal suffering from various losses. Although this requires the assumption that all the losses act on the signal coherently, thereby no nontrivial physics emerges and one can map the whole setup with a target having an effective lower reflectivity. In the near future, it can be interesting to find whether the effects of such losses on the performance of the illumination process are \textcolor{black}{drastic or not}.  We believe that our analysis provides a consistent way to analyze quantum illumination with non-Gaussian states, especially when imperfections are affecting the process. 
	Moreover, our work provides an appropriate platform for classifying Gaussian and non-Gaussian states based on their performance in quantum illumination.

	\section{Acknowledgement}
	
	RG, SR, and ASD acknowledge the support from Interdisciplinary Cyber Physical Systems (ICPS) program of the Department of Science and Technology (DST), India, Grant No.: DST/ICPS/QuST/Theme- 1/2019/23. TD acknowledges support by Foundation for Polish Science (FNP), IRAP project ICTQT, contract no. 2018/MAB/5, co-financed by EU Smart Growth Operational Programme. 
	We  acknowledge the uses of \href{http://arma.sourceforge.net/docs.html}{Armadillo} -- a high quality linear algebra library for the C++ language \cite{arma1,arma2},  \href{https://github.com/titaschanda/QIClib}{QIClib} -- a modern C++ library for general purpose quantum information processing and quantum computing  \cite{qiclib} and  cluster computing facility at Harish-Chandra Research Institute.

\section*{Supplementary material}

 	\section{Non-Gaussian operations: Adding and subtracting photons }
	\label{sec:app1}
	\textcolor{black}{	Non-Gaussianity has been proven to be a performance enhancer in different contexts, ranging from entanglement \cite{Cerfaddsub,tamo2016} to non-locality \cite{SappyTamoGBell}, and in the case of QI as well \cite{NGIllu}. Among a plethora of de-Gaussification techniques, engineering non-Gaussian state via photon addition and subtraction offers the advantage of experimental realizability \cite{trepsexp,multilight}  which, in turn, motivates us to take this route of de-Gaussification. By “photon addition/subtraction”,  we mean  application of the bosonic creation $\hat{a}^\dagger$ and annihilation operators $\hat{a}$, which are most commonly used in the field of quantum optics involving the non-Gaussian operations, (cf. \cite{Cerfaddsub,tamo2016,YangPAPS,SappyTamoGBell,PAPS,PAPS2}). We also note that bare raising and lowering operators, $E^+ ~ \text{and} ~ E^-$, might also be the choice, but the effect will remain the same after normalizing the resulting states.\\
	Before moving on to non-Gaussian resources, let us first consider the Gaussian two-mode squeezed vacuum
	state, given by 
	\begin{eqnarray}
	|\psi_r\rangle = \sum_{n=0}^\infty c_n |n,n\rangle,   
	\label{eq:tmsv}
	\end{eqnarray}
	where $c_n=(1-x)^{\frac{1}{2}} x^{\frac{n}{2}}$ with $x=\tanh^2 r$, $r$ being the squeezing parameter and $\lbrace |n\rangle \rbrace$ representing the Fock basis.	
	\\
	By de-Gaussifying the TMSV state  in both the modes (i.e., by adding  \(k\) and \(l\) photons  in the first and the second mode respectively),  the (normalized) photon-added  state   can be represented as
	\cite{Cerfaddsub}
	\begin{eqnarray}
	|\psi_r^{(k,l)}\rangle = \sum_{n=0}^\infty c_n^{(k,l)}|n+k,n+l\rangle,  
	\label{eq:added_state}
	\end{eqnarray}
	where
	\begin{eqnarray}
	c_n^{(k,l)} = \frac{x^{\frac{n}{2}}}{\sqrt{{_2}F_1(k+1,l+1,1,x)}} \sqrt{\binom{n+k}{k}\binom{n+l}{l}},\nonumber \\
	\label{eq:patmsv}
	\end{eqnarray}
	${_2}F_1$ is the Gauss Hypergeometric function, and $c_n^{(0,0)} = c_n$,  while  the photon-subtracted state obtained after subtracting photons from both the modes can be written as
	\begin{eqnarray}
	|\psi_r^{(-k,-l)}\rangle = \sum_{n=\max \{k,l\}}^\infty c_n^{(-k,-l)}|n-k,n-l\rangle  
	\label{eq:sub_state}
	\end{eqnarray}
	with
	\begin{eqnarray}
	c_n^{(-k,-l)} = \frac{x^{\frac{n-k}{2}}}{\sqrt{{_2}F_1(k+1,k+1,1+k-l,x)}} \sqrt{\frac{\binom{n}{k}\binom{n}{l}}{\binom{k}{l}}}. \nonumber \\
	\label{eq:pstmsv}
	\end{eqnarray}
	Without loss of generality, here we assume $k \geq l$. Notice, however, that Eq. (\ref{eq:pstmsv}) holds even for \(l > k\) with \(k\) and \(l\) being interchanged  in Gauss Hypergeometric function and in the denominator. \\
	\textbf{Note.} Instead of two modes, if addition (subtraction) is performed in a single mode, say, in the second mode,   the corresponding output state can be obtained by  putting \(k=0\), i.e., 
	\begin{eqnarray}
	c_n^{(0,l)} = x^{\frac{n}{2}}(1&-&x)^{\frac{1+l}{2}} \sqrt{\binom{n+l}{l}},  
	\label{spatmsv}
	\end{eqnarray}
	and in the case of subtraction, it is
	\begin{eqnarray}
	c_n^{(0,-l)} &=& x^{\frac{n-l}{2}}(1-x)^{\frac{1+l}{2}}\sqrt{\binom{n}{l}}.
	\label{eq:singlemode_cn's}
	\end{eqnarray}
	Notice that  here we consider the subtraction operation beyond \(k \geq l\). }
	
	\section{The classical limit}
\label{sec:app2}
\textcolor{black}{	The scheme \cite{QvCoherent,Bound1,TrepsArxiv,GIllu} which uses coherent states, $|\sqrt{N_S}~\rangle$, as a signal probe for illumination can be referred to as  classical illumination method. The  Chernoff bound for the same with one copy of the probe state is computed to be
	\begin{eqnarray}
	\mathcal{Q}_c = e^{- \kappa N_S (\sqrt{N_B} - \sqrt{N_B + 1})^2},
	\label{eq:classical1}
	\end{eqnarray}
	which in the limit $N_B >> 1$ reduces to
	\begin{eqnarray}
	\mathcal{Q}_c \approx e^{-\frac{\kappa N_S}{4N_B}}.
	\end{eqnarray}
	The corresponding minimum error probability for $M$-copies of the coherent state is then upper bounded by 
	\begin{eqnarray}
	P_M^c \leq \frac{1}{2}\mathcal{Q}_c^M.
	\end{eqnarray}
	In this paper, when we observe $P_M < P_M^c$, where the average number of photons in the signal probe is fixed to $N_S$ in both the cases under comparison, we claim that quantum advantage is obtained in the QI method. }
	
		\section{Method for computing Chernoff bounds}
	\label{app:numerical}
	\textcolor{black}{	Let us discuss the method which we use to calculate the Chernoff bounds for non-Gaussian states. 
	The density matrices corresponding to $\rho_1$ and $\rho_0$  are constructed according to Eqs. ($2$) and ($5$) of the main text . The truncated infinity, i.e., $N$ in the summation is chosen in such a way, that the precision in trace and the computed CB is less than $10^{-8}$. This implies that the numerically calculated values of trace and the CB are correct up to 8 decimal points i.e., if we increase our truncated $N$ to a new $N'$, where $N' > N$,  the value up to the eighth decimal point remains unaltered. Under such conditions, the matrices have dimension $(N \pm k+1)(2N \pm l+1) \times (N \pm k+1)(2N \pm l+1)$, where $(N \pm k +1) \times (N \pm k+1)$  denotes the size of the idler subsystem and $(2N \pm l+1) \times (2N \pm l+1)$ is that for the signal subsystem, with $+ \: \text{and} \: -$ representing the photon addition and subtraction respectively. In the case of the TMSV state, and the photon-added states, it  is observed that any $N \geq 35$ is sufficient while for photon subtraction,  we have to take $N \geq 45$. In the case of  mixed states convergence is achieved for the same limits. Thereafter, we optimize over $\alpha$ to obtain the Chernoff bound. All the calculations are performed for a fixed value of $\kappa$, $N_B$, and $x$ with $k$ and $l$ running from $0$ to $10$. In most of our analysis,  we set $\kappa = 0.01$, representing the target reflectivity, unless mentioned otherwise and the mean number of photons in the thermal background is taken to be  unity. A major part of our calculations are done by setting  $x = 0.2 \: \: \text{and} \: \: 0.05$.\\
	\textbf{Remark 1.} An error probability of $0.5$ is trivial and any deviation from the same may be considered as an enhancement of performance. In the entirety of our analysis, we consider any difference which is $\leq 10^{-4}$ to be negligible, and thus a Chernoff bound $\geq 0.4999$ is rounded off to  be $0.5$. This turns out to be a necessary assumption since we find that if a vacuum state, i.e., nothing impinges on the beam splitter of reflectivity $\kappa = 0.01$, and the average photon number in the background thermal noise is taken to be $\bar{n}$ = 1.0,  the CB is equal to $0.49997$, which is less than 0.5 up to the fifth decimal point. A detailed calculation is given in Sec. \ref{appsec:CBvac}. From this calculation, it is plausible to assume that r the Chernoff bound between the range (0.49997, 0.5) may not be required. Hence, we can safely claim that $0.49997$ is our $0.5$. Moreover, for the CB which is less than $0.49997$,  the numerical precision reported here is up to the eighth decimal place.  Throughout our work, we represent states with photons added in one mode as ``sPA" and those with photons added in both modes as ``PA" while photon-subtracted states are dubbed as  ``PS". Interestingly, states with photons added in the signal mode behave exactly similar to those with subtracted photons from the idler mode and vice versa \cite{SappyTamoGBell}. Thus, we deal only with single mode addition of photons and the performance of probes with subtracted photons from one mode follows trivially.\\
	\textcolor{black}{\textbf{Remark 2}.  Let us stress again the reasoning behind considering the tolerance mentioned in the paper. In this work, we have considered the Chernoff bound for the illumination protocol with one copy of the probe state because the multi-shot results follow easily due to the bound being asymptotically tight. Hence, it is enough to calculate analytically the Chernoff bound for a single copy level (Q), as we are interested in obtaining the probability of error for M-copies of the probe state, and it follows the relation $P_M <= 0.5(2Q)^M$. We plot all the figures with a tolerance value $10^{-14}$.
}
}

\section{Quantum advantage by varying target reflectivity}
\label{sec:app5}	
\begin{figure}
		\centering
		\includegraphics[width=\linewidth]{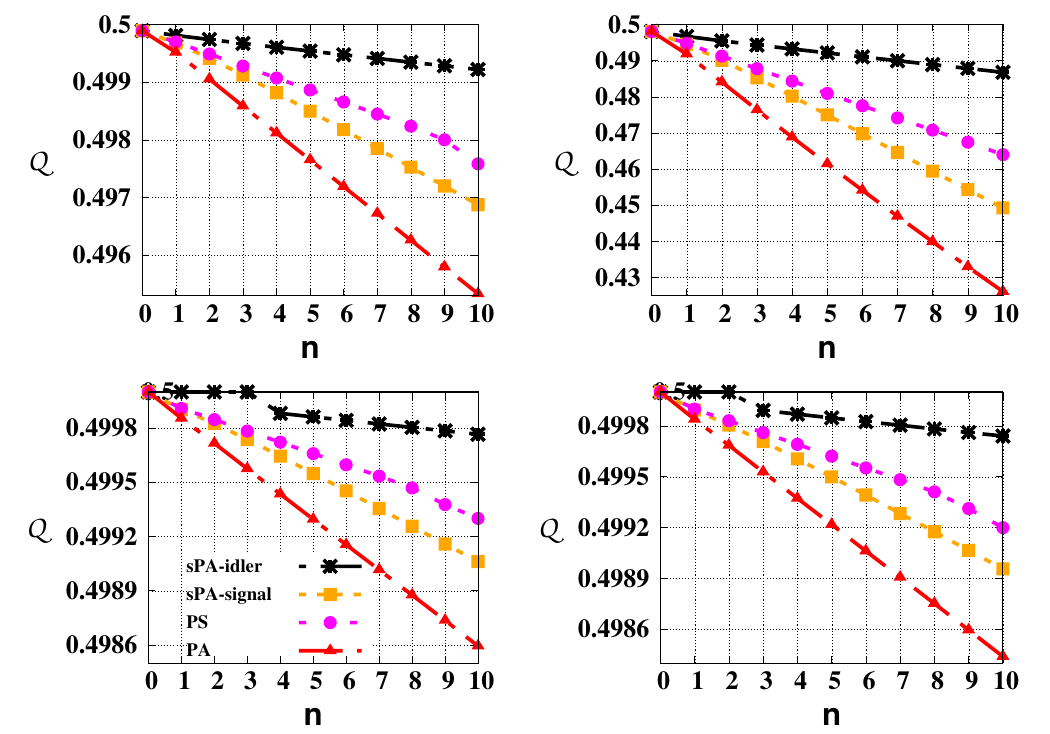}
		\caption{(Color Online.) Performance of Chernoff bound (\(y\)-axis) with respect to \(n\)  (\(x\)-axis) by varying reflectivity, \(\kappa\). (Upper panel)  $\kappa = 0.0009$ on the left and $\kappa = 0.001$ on the right. (Lower panel, left) $\kappa = 0.003$  and (right) $\kappa = 0.05$.  The  TMSV state from which non-Gaussian states are created has $x = 0.2$. 
			All other specifications are the same as in Fig. ($2$) of the main text.
			Both axes are dimensionless.	 }
		\label{ref_fig}
	\end{figure}

\textcolor{black}{	In QI protocol a weakly reflecting target is modeled by a beam splitter of low reflectivity. Now, the detection scheme may have to cater to various targets having different reflectivities $\kappa$. If  $\kappa$ is low,  the efficiency of the protocol always decreases. 
	In this context,  it is interesting  to find the trade-off between the  low  reflectivity
	and the enhancement of QI due to non-Gaussianity.
	In the case of the  TMSV state, we find that for $\kappa \leq 0.001$, there is a negligible advantage over the blind guesses of the target,  while non-Gaussian states can still provide  a better detection probability \(\mathcal{Q}\) even for such a low reflectivity as illustrated in Fig. \ref{ref_fig}.\\
	For  low reflectivity, even the state with photons added only in the idler mode is unable to provide low CB unless the  added number of photons is  $n \geq 4$. However, similar to previous scenarios, single photon addition (subtraction) from the signal (idler) mode incorporates enough non-Gaussianity to have a low value of CB. 
	The hierarchy of the non-Gaussian states in terms of lower error probability is maintained such that photon-subtracted states perform better than states with photons added in the idler but cannot overpower states with photons added in the signal mode. }
	
	\begin{figure}[ht]
		\centering
		\includegraphics[width=\linewidth]{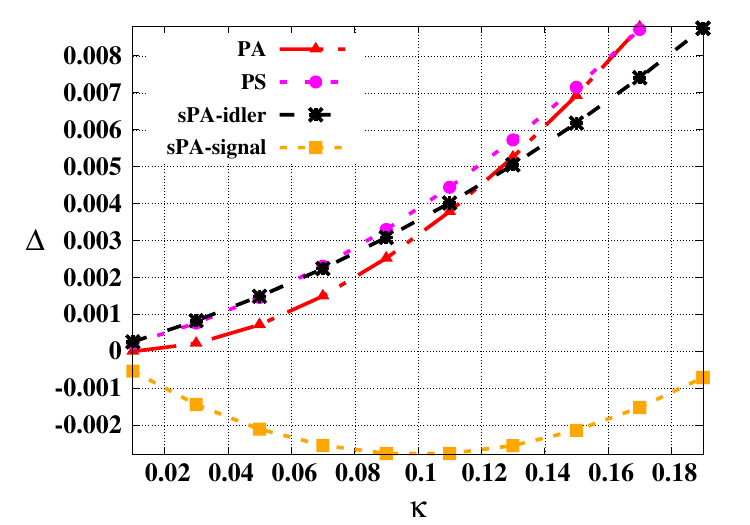}
		\caption{(Color Online.) $\Delta$ (ordinate) vs. $\kappa$ (abscissa). Here $n= 5$. All other specifications are the same as in Fig. ($2$) of the main text. Both axes are dimensionless.}
		\label{cbound_ref_Ns}
	\end{figure}		
\textcolor{black}{	\textit{Effect of varying target reflectivity on quantum efficiency.}
	For a fixed number of photons added (subtracted), we now compare \(\Delta\) obtained from different non-Gaussian states by varying \(\kappa\). 
	For reflectivities of the order of $10^{-2}$ or higher,  the addition of photons in the idler mode, and in both the modes as well as the subtraction of photons from both the modes can outperform the classical protocol, thereby demonstrating the quantum advantage (see Fig. \ref{cbound_ref_Ns} where the number of photons added (subtracted) in (from) a single or both the modes is 5).
	However, the states having photons added in the signal mode (subtracted from the idler mode) can never give $\Delta > 0$, and thus, it is not suitable as a probe state.
	Notice, however, that for extremely low reflectivity, $\kappa < 10^{-2}$, neither the TMSV state nor the non-Gaussian states can provide any quantum advantage, thereby showing the coherent state-based protocol to be good enough. \\
	\textbf{Note.} For $\kappa < 0.01$, the Chernoff bound for the  TMSV state gives a difference from $0.5$ in the fifth decimal place (after $0.4999$) only for small $N_S$ i.e., up to a squeezing of $x = 0.4$ or lower. In this section, we deal with non-Gaussian states whose photon number $N_S$ is much higher than that of the TMSV state having $x = 0.4$. Thus for a TMSV state having a photon number equal to the non-Gaussian states, the Chernoff bound is indeed below 0.5, and our approximation  does not lead to any underestimated results. }
	
		\section{Quantum illumination with imperfect photon addition or subtraction}
	\label{app:imper}
	\textcolor{black}{Due to several imperfections, e.g.,  dark counts of the detector \cite{dark1,dark2,dark3}, the probe states produced may not have the desired number of added or subtracted photons. As a result, the final state is a mixture of states with varying levels of non-Gaussianity. We assume that, for a given number $k$ of photons to be added or subtracted, the state is mixed with other states having $k, k-1, k-2, \ldots, k-m$ (\(m \leq k\)) added or subtracted photons, with different probabilities. Here, $m$ represents the cutoff in the discrepancy which can be incorporated due to the imperfect creation process. The state, therefore, takes the form as
	\begin{equation}
	\tilde{\rho}_{\pm k} = \sum_{i = 0}^{m} p_i \rho_{\pm |k - i|}.
	\label{imperfect_eq}
	\end{equation}
	We will be interested to compute the performance of QI using $\tilde{\rho}_{\pm k}$ and compare it with the corresponding classical bound.\\
	Let us demonstrate the case  when the photonic operations are imperfect, and non-Gaussianity still continues to provide improvements  in QI.
	To demonstrate it, we consider imperfect subtraction of photons with $k = l = 2$ and $m \leq 2$. Thus we have a two photon-subtracted state, mixed with a single photon-subtracted state and the Gaussian TMSV state as the probe in the illumination scheme which reads as
	\begin{equation}
	\tilde{\rho} = p \rho_2 + p' \rho_1 + p'' \rho_{TMSV}
	\label{imperfect_eq_state}
	\end{equation}
	with $p + p' + p'' = 1$. Here $\rho_i$ represents the state  where $i$th number of photons have been subtracted from both signal and the idler modes.}
	
	\begin{figure}[h]
		\centering
		\includegraphics[width=\linewidth]{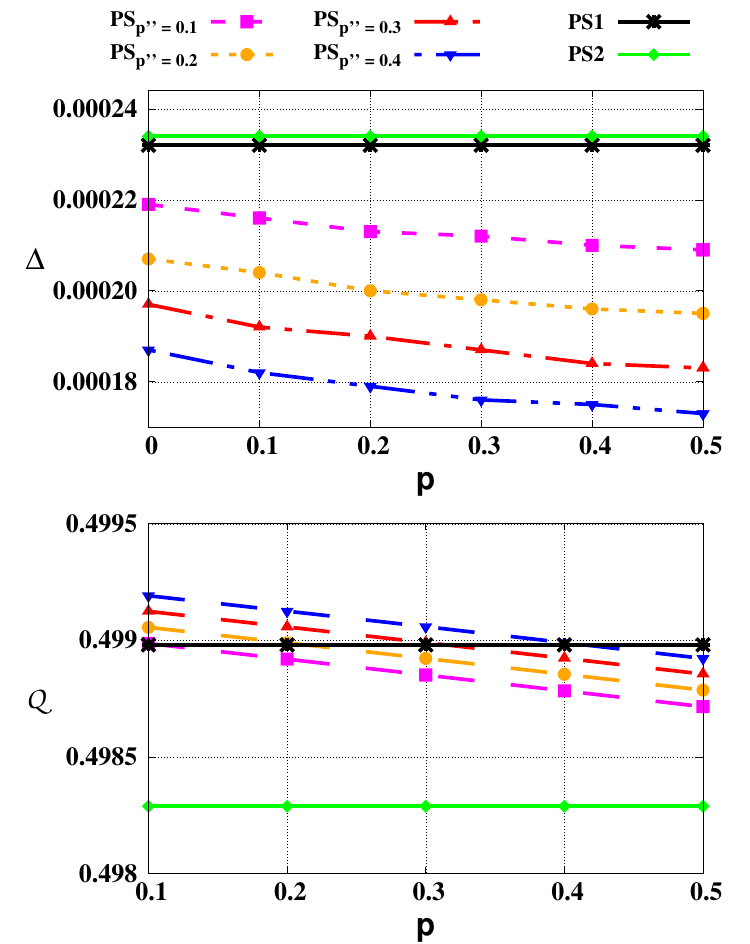}
		\caption{(Color Online.) Consequence of imperfect photon subtraction device on illumination.  (Upper panel) Quantum advantage ($\Delta$) (ordinate) against $p$  (abscissa) while (lower panel)  $\mathcal{Q}$  vs. \(p\). Stars  and diamonds represent the states without imperfection, i.e.,  photon-subtracted state from a single mode and  subtraction of photons from both  modes respectively. The imperfect states given in Eq. (\ref{imperfect_eq_state}) are for $p'' = 0.1$ (squares), $p'' = 0.2$ (upward triangles), $p'' = 0.3$ (circles) and $p'' = 0.4$ (downward triangles). Both axes are dimensionless.}
		\label{ineff_fig_Ns}
	\end{figure}

\textcolor{black}{	It is seen that, even with imperfect probe generation, the CB is always lower than that with the Gaussian state, i.e., \(\mathcal{Q} (\tilde{\rho}) > \mathcal{Q} (\rho_{TMSV}) \). 
	We consider three probabilities, $p$--corresponding to the perfect two photon-subtracted states, $p'$--for the single photon-subtracted state, and $p''$--for the TMSV state respectively. For our investigation, we set a fixed value of $p''$ and vary $p$, with $p'$ being determined by the normalization condition in Eq. \eqref{imperfect_eq_state}. As the probability of obtaining the required state decreases, the CB monotonically increases, as is evident from the bottom panel in Fig. \ref{ineff_fig_Ns}.  Moreover, we notice that 
	as long as the desired perfect state has $p \geq p''$, the CB possesses a lower value than that obtained for the photon-added state in a single mode. 
	Therefore,  even if the required non-Gaussian state is obtained with a very small probability, its mere presence is enough to ensure a low error probability for discriminating targets. \\
	\textit{Deterioration of the performance of QI under imperfection. }
	Comparing the performance of non-Gaussian states with that of a coherent state having the same signal strength, we find that even if the photon subtraction mechanism is imperfect, the resulting state can always beat the classical limit, as depicted in the top panel of Fig. \ref{ineff_fig_Ns} by the positive values of $\Delta$. However, such states can never perform better than the perfect two-photon- and single-photon-subtracted states. As the probability, $p''$ of generating the TMSV state,  due to the apparatus's imperfection, the illumination efficiency falls considerably, thereby implying that even a highly imperfect photon subtraction protocol can yield substantial quantum advantage. Our results indicate that photon subtraction is an efficient process to obtain a quantum advantage over the coherent state even if the state generation process is not perfectly optimal. Notice that in the case of photon addition, such an advantage cannot be seen. \\
	\textcolor{black}{\textbf{Remark:} An important observation that we wish to discuss here, is the asymmetry in performance between the photon-added and the photon-subtracted states in the illumination protocol. The asymmetry in performance between the photon-added and photon-subtracted states is observed in several other quantum information protocols like teleportation \cite{DellAnno_PRA_2007,DellAnno_PRA_2010,DellAnno_PRA_2010_2,Patra_arxiv_2022}. However, the reason behind this is still not clear in the literature. \\
Although the single mode photon-added and subtracted display the same performance in illumination, the  states have different levels of non-Gaussianity, as has been demonstrated in \cite{Roy_PRA_2020, Benlloch_PRA_2012}. It is true that in the illumination protocol, the idler channel needs to be retained for a period of time when it can undergo loss. Still, to our knowledge, this is not the reason behind the difference in Chernoff bound between the photon-added and subtracted states. This is because the idler subsystem of both the photon-added and photon-subtracted states would suffer the same amount of noise and thus their performance would be equally affected. Modeling gain in the idler channel should intuitively help the illumination process, but still does not explain the apparent asymmetry, since the gain too would aid both states by an equal measure. It was shown in \cite{Benlloch_PRA_2012} that states with an unequal number of photons added and subtracted from corresponding modes have different entanglement and such states do perform differently in the illumination protocol, case in point: the signal photon-added (idler photon-subtracted) and idler photon-added (signal photon-subtracted) states. However, in terms of entanglement profile too, the states with an equal number of photons added and subtracted from both the modes, are exactly alike as was demonstrated in \cite{NGIllu,Benlloch_PRA_2012}. Thus, even the entanglement structure does not help in explaining the asymmetry in performance between the states having an equal number of photons added or subtracted from two modes. It should be noted that only non-Gaussian states with an equal number of photons added or subtracted from both modes show this apparent asymmetry despite having the same entanglement content. We leave this as an open question along with the idea of modeling gain in the idler channel.}}
	
	\section{Chernoff Bound for vacuum state $\lambda = 0$}
	\label{appsec:CBvac}
 
	When $\lambda = 0$, the TMSV state is just $\ket{00}_{IS}$. Hence when the target is absent, we have
	\begin{equation}
	\rho_0 = \ket{0}\bra{0}_{I} \otimes \rho_T  = \sum_n \frac{\bar{n}^n}{(1 + \bar{n})^{n+1}} \ket{0}\bra{0}_{I} \otimes \ket{n}\bra{n}_T
	\end{equation}
	where $\rho_T = \sum_n \frac{\bar{n}^n}{(1 + \bar{n})^{n+1}}\ket{n}\bra{n}_T$ is the thermal state, for which we consider $\bar{n} = 1$.
	On the other hand, when the target is present we obtain
	\begin{eqnarray}
	 && \nonumber U_{ST}(\rho_{IS}\otimes \rho_T) U^\dagger_{ST} \\
	 && \nonumber =   U_{ST} \left(\ket{0}\bra{0}_{IS} \otimes \sum_n \frac{\bar{n}^n}{(1 + \bar{n})^{n+1}}\ket{n}\bra{n}_T \right) U^\dagger_{ST} \\
	&& \nonumber =  U_{ST} \Big(\ket{0}\bra{0}_{I} \otimes \sum_n \frac{\bar{n}^n}{(1 + \bar{n})^{n+1}} \frac{1}{n!} \times \\
       && ~~~~~~~~ (\hat{a}^\dagger_T)^n \ket{00}\bra{00}_{ST}\hat{a}_T^n  \Big) U^\dagger_{ST} \label{rho_1_app},
	\end{eqnarray}
	where $U_{ST}$ is the unitary representing the beam splitter which acts as the target and $\hat{a}_T$ is the annihilation operator corresponding to the input thermal mode.
	Now we know that $U^\dagger_{ST} \hat{a}_T U^\dagger_{ST} = \sqrt{\tau} \hat{a}_S' + \sqrt{\kappa} \hat{a}_T'$ with $\tau$ being the target transmissivity and $\kappa$ is the target reflectivity. $\hat{a}_S' ~ \text{and} ~ \hat{a}_T'$ are the annihilation operators for the output signal and thermal modes. Upon substituting this in Eq. \eqref{rho_1_app}, it reads as
	\begin{eqnarray}
	\nonumber &&  U_{ST}(\rho_{IS}\otimes \rho_T) U^\dagger_{ST} \\
	\nonumber &=& \ket{0}\bra{0}_{I} \otimes \sum_n \frac{\bar{n}^n}{(1 + \bar{n})^{n+1}} \frac{1}{n!} (\sqrt{\tau} \hat{a}_S'^\dagger + \sqrt{\kappa} \hat{a}_T'^\dagger)^n \times \\
 \nonumber && \hspace{5em} \ket{00}\bra{00}_{ST} (\sqrt{\tau} \hat{a}_S' + \sqrt{\kappa} \hat{a}_T')^n \\
	\nonumber &=& \nonumber \ket{0}\bra{0}_{I} \otimes \sum_n \frac{\bar{n}^n}{(1 + \bar{n})^{n+1}} \frac{1}{n!} \sum_{r,r'} \binom{n}{r} \binom{n}{r'} \times \\
    \nonumber  &&\hspace{-3em} (\sqrt{\tau})^{2n - r - r'}(\sqrt{\kappa})^{ r + r'} (\hat{a}_S'^\dagger)^{n-r} (\hat{a}_T'^\dagger)^r \ket{00}\bra{00}_{ST} (\hat{a}_S')^{n-r'} (\hat{a}_T')^{r'} \\
\nonumber	&=& \nonumber \ket{0}\bra{0}_{I} \otimes \sum_n \frac{\bar{n}^n}{(1 + \bar{n})^{n+1}} \frac{1}{n!} \sum_{r,r'} \binom{n}{r} \binom{n}{r'} \times \\
\nonumber && \hspace{-3em} (\sqrt{\tau})^{2n - r - r'}(\sqrt{\kappa})^{ r + r'} \sqrt{(n-r)!r!} \sqrt{(n-r')!r'!} \times \\
 && \hspace{5em} \ket{n-r,r}\bra{n-r',r'}_{S'T'} 
	\end{eqnarray}
	Here $S, T$ denote the input signal and thermal modes respectively, while $S', T'$ stand for the corresponding output modes. Taking trace over $T'$ we obtain, 
	\begin{eqnarray}
	\nonumber && \rho_1 = tr_{T'} (U_{ST}(\rho_{IS} \otimes \rho_T) U^\dagger_{ST}) \\
	&=& \nonumber  \ket{0}\bra{0}_{I} \otimes \sum_n \frac{\bar{n}^n}{(1 + \bar{n})^{n+1}} \frac{1}{n!} \binom{n}{r}^2  (\sqrt{\tau})^{2(n - r) } \times \\
                  && \nonumber (\sqrt{\kappa})^{2 r} (n-r)!r! \ket{n-r}\bra{n-r}_{S'} \\
	&=& \nonumber \ket{0}\bra{0}_{I} \otimes \sum_{n = 0}^\infty \frac{\bar{n}^n}{(1 + \bar{n})^{n+1}}  \sum_{r = 0}^n \binom{n}{r}  \tau^{r}\kappa^{n- r }  \ket{r}\bra{r}_{S'} \\
	&=& \nonumber \ket{0}\bra{0}_{I} \otimes \sum_{r = 0}^\infty \left( \sum_{n = r}^\infty \frac{\bar{n}^n}{(1 + \bar{n})^{n+1}}  \binom{n}{r}  \tau^{ r}\kappa^{n-r} \right) \ket{r}\bra{r}_{S'} \\
	&=&  \ket{0}\bra{0}_{I} \otimes \sum_{r = 0}^\infty \frac{(1-\kappa)^r \bar{n}^r}{(1 + \bar{n} (1-\kappa) )^{r+1}} \ket{r}\bra{r}_{S'}
	\end{eqnarray}
	
	By ignoring $\ket{0}\bra{0}_{I}$ the Chernoff bound can be calculated as
	\begin{eqnarray}
	&& \nonumber \frac 12 \min_{0 \leq \alpha \leq 1} Tr [\rho_0^\alpha \rho_1^{1-\alpha}]  \\
	&& \hspace{-3em} = \frac 12 \min_{0 \leq \alpha \leq 1} \sum_{n=0}^\infty \left( \frac{\bar{n}^n}{(1 + \bar{n})^{n+1}}\right)^\alpha \left(\frac{(1-\kappa)^n \bar{n}^n}{(1 + \bar{n} (1 - \kappa) )^{n+1}} \right)^{1 - \alpha} ~~~~~~
	\end{eqnarray}
	For $\bar{n} = 1$, and $\kappa = 0.01$, the minimum value is $0.49997$. From this calculation, it is plausible to assume that  the Chernoff bound between the range $(0.49997, 0.5)$ may not be required. Hence, we can safely claim that $0.49997$ is our $0.5$. Note, however, that the CB which is $< 0.49997$ for $\kappa = 0.01$ and $\bar{n} = 1.0$ is considered up to $8$ decimal points.\\

	\bibliographystyle{elsarticle-num-names}
	\bibliography{bib}

\end{document}